    \documentclass[
        fleqn,
        usenatbib,
    ]{mnras}
    \usepackage{
        amsmath,
        amssymb,
        newtxtext,
        newtxmath,
        graphicx,
        ae, aecompl,
        booktabs,
        xcolor,
    }
    \usepackage[T1]{fontenc}

    \newcommand*{\rd}[2]{\frac{\mathrm{d}#1}{\mathrm{d}#2}}
    
    \newcommand*{\rdil}[2]{\mathrm{d}#1 / \mathrm{d}#2}

    \newcommand*{\bm}[1]{\boldsymbol{\mathbf{#1}}}

    \newcommand*{\abs}[1]{\left|#1\right|}
    
    \newcommand*{\p}[1]{\left(#1\right)}
    \newcommand*{\s}[1]{\left[#1\right]}
    
    \colorlet{Corr}{red}
    \newlength{\colummwidth}
    \DeclareMathOperator*{\sgn}{sgn}

\title[Eccentric Dynamical Tides]{Dynamical Tides in Eccentric Binaries
Containing Massive Main-Sequence Stars: Analytical Expressions}
\author[Y. Su, D. Lai.]{
Yubo Su$^1$,
Dong Lai$^1$\\
$^1$ Cornell Center for Astrophysics and Planetary Science, Department of
Astronomy, Cornell University, Ithaca, NY 14853, USA
}

\date{Accepted XXX\@. Received YYY\@; in original form ZZZ}

\pubyear{2020}

\begin{document}\label{firstpage}
\pagerange{\pageref{firstpage}--\pageref{lastpage}}
\maketitle

\begin{abstract}
    Tidal evolution of eccentric binary systems containing at least one massive
    main-sequence (MS) star plays an important role in the formation scenarios
    of merging compact-object binaries.  The dominant dissipation mechanism in
    such systems involves tidal excitation of outgoing internal gravity waves at
    the convective-radiative boundary and dissipation of the waves at the
    stellar envelope/surface. We have derived analytical expressions for the
    tidal torque and tidal energy transfer rate in such binaries for arbitrary
    orbital eccentricities and stellar rotation rates. These expressions can be
    used to study the spin and orbital evolution of eccentric binaries
    containing massive MS stars, such as the progenitors of merging neutron star
    binaries. Applying our results to the PSR J0045-7319 system, which has a
    massive B-star companion and an observed, rapidly decaying orbit, we find
    that for the standard radius of convective core based on non-rotating
    stellar models, the B-star must have a significant retrograde and
    differential rotation in order to explain the observed orbital decay rate.
    Alternatively, we suggest that the convective core may be larger as a result
    of rapid stellar rotation and/or mass transfer to the B-star in the recent
    past during the post-MS evolution of the pulsar progenitor.
\end{abstract}


\section{Introduction}

The physics of tidal dissipation in massive, main-sequence (MS) stars (i.e.\
having a convective core and radiative envelope) under the gravitational
influence of a companion was first studied by \citet{zahn1975dynamical}
\citep[see also][]{savonije1983tidal, goldreich1989tidal}. The dominant
dissipation mechanism is through the \emph{dynamical tide}, in which the
time-dependent tidal potential of the companion excites internal gravity waves
(IGWs) at the convective-radiative boundary (RCB). As the wave propagates
towards the surface, its amplitude grows, and the wave dissipates efficiently
\citep{zahn1975dynamical, goldreich1989tidal, su2020}. The contribution due to
viscous dissipation in the convective core is expected to be subdominant.

The original expression describing the torque due to dynamical tides by
\citet{zahn1975dynamical} is very sensitive to the global properties of the
star. \citet{kushnir} present an updated derivation of the tidal torque
that depends only on the local stellar properties near the RCB, eliminating many
uncertainties from the Zahn's original expression. In
\citet{zaldarriaga2018expected}, the authors use the new expression to study
tidal synchronization in binaries consisting of a Wolf-Rayet star and a black
hole, the likely progenitors to merging black-hole binaries observed by
LIGO/VIRGO\@.

These previous works all apply to nearly circular binaries. However, massive
stars can often be found in high-eccentricity (high-e) systems, such as binaries
consisting of one MS star and one neutron star (NS). The NS is formed with a
large kick velocity \citep[e.g.][]{lai2001pulsar, janka2021supernova}, giving
rise to a high-e binary. Several such high-e MS-NS systems have been discovered
\citep[e.g.][]{kaspi1994massive, johnston1994radio, champion2008eccentric}.
These are the progenitors of double NS systems \citep[e.g.][]{tauris2017}. An
important issue is to understand whether such high-e systems can circularize
prior to mass transfer or a common envelope phase \citep{vigna2020common,
vicklai2021}.

The purpose of this paper is to derive easy-to-use, analytical expressions for
the effects of dynamical tides for high-e binaries with massive MS stellar
companions. In Section~\ref{s:background}, we summarize the equations of
dynamical tides involving IGWs in circular binaries and existing techniques for
studying high-e systems. In Section~\ref{ss:objective}, we evaluate the effect
of dynamical tides in high-e systems containing massive MS stellar companions,
including the torque and orbital decay rate. In Section~\ref{s:j00457319}, we
apply our results to the pulsar-MS binary J0045-7319, for which a non-zero
orbital decay rate has been measured. Finally, we summarize our results and
discuss the uncertainties in Section~\ref{s:disc}.

\section{Dynamical Tides in Massive Stars}\label{s:background}

\subsection{Circular Binaries}\label{ss:2_circ}

We first review the case where the binary is circular. Let $M$ be the mass of
the MS star, $M_2$ the mass of the companion, $a$ the semimajor axis of the
binary, and $\Omega$ the angular frequency (mean motion) of the binary. The
tidal torque exerted on the star by the companion due to tidal excitation of
IGWs at the RCB is \citep{kushnir}
\begin{align}
    T_{\rm circ}(\omega) ={}& T_0 \sgn\p{\omega}
        \Big|\,\frac{\omega}{\Omega}\,\Big|^{8/3} \label{eq:kushnir_torque},
\end{align}
where
\begin{align}
    T_0 \equiv{}& \beta_2\frac{GM_2^2r_{\rm c}^5}{a^6}
            \p{\frac{\Omega}{\sqrt{GM_{\rm c}/r_{\rm c}^3}}}^{8/3}
            \frac{\rho_{\rm c}}{\bar{\rho}_{\rm c}} \p{1 - \frac{\rho_{\rm
            c}}{\bar{\rho}_{\rm c}}}^2,\\
    \beta_2 \equiv{}&
        \s{\frac{r_{\rm c}}{g_{\rm c}}
            \p{\rd{N^2}{\ln r}}_{r = r_{\rm c}}}^{-1/3}
                \s{\frac{3^2\Gamma^2(1/3)}{40\pi 12^{2/3}} \alpha^2}.
\end{align}
Here, $\omega \equiv 2\Omega - 2\Omega_{\rm s}$ is the tidal forcing frequency,
$\Omega_{\rm s}$ is the spin of the MS star, $N$ is the Br\"unt-Vaisala
frequency, $r$ is the radial coordinate within the star, and $r_{\rm c}$,
$M_{\rm c}$, $g_{\rm c}$, $\rho_{\rm c}$, and $\bar{\rho}_{\rm c}$ are the
radius of the RCB, the mass contained within the convective core, the
gravitational acceleration at the RCB, the stellar density at the RCB, and the
average density of the convective core respectively. $\Gamma$ is the gamma
function, $\alpha$ is a numerical
constant of order unity given by Eq.~(A32) of \citet{kushnir}, and $\beta_2
\approx 1$ for a large range of stellar models (Fig.~2 of \citealp{kushnir}). In
Eq.~\eqref{eq:kushnir_torque}, we have expressed the various factors such that
$T_0$ contains all the spin-independent terms.

The above result (Eq.~\ref{eq:kushnir_torque}) assumes that tidally excited IGWs
dissipate completely as they propagate outwards towards the stellar surface.
Such dissipation can happen either through radiative damping or nonlinear
effects. Recent hydrodynamical simulations of the IGW breaking process in the
stellar envelope \citep{su2020} show that the nonlinear damping of the
outward-propagating IGWs is due to the development of a narrow critical layer.
This critical layer divides the star into a asynchronously-rotating interior and
a synchronously-rotating exterior \citep[see][]{goldreich1989tidal}, and it
efficiently absorbs the angular momentum of the incident IGWs ($\sim 70\%$ of
incident flux, \citealp{su2020}). If the IGWs instead reflect before dissipating
completely, then standing waves are set up in the stellar interior. These
internal oscillations then dissipate due to radiative damping and nonlinear mode
couplings (for the former, see e.g.\ \citealp{lai1996, lai1997,
kumar1997differential, kumar1998}, and for the latter, see e.g.\
\citealp{oleary}). Except when a tidal forcing frequency is resonant with a
stellar oscillation mode, the traveling-wave assumption represents an upper
bound on the tidal torque (see e.g.\ \citealp{yu2020non, yu2021tidally}
concerning tidal dissipation in white dwarf binaries).

Note additionally that we use Eq.~\eqref{eq:kushnir_torque} from \citet{kushnir}
instead of the classic expression from \citet{zahn1975dynamical}, which is given
by:
\begin{equation}
    T_{\rm circ}^{\rm (Zahn)}(\omega) = \frac{3}{2}\frac{GM_2^2 R^5}{a^6}
        E_2\p{\frac{\omega}{\sqrt{GM/R^3}}}^{8/3},
\end{equation}
where $M$ and $R$ are the mass and radius of the MS star and $E_2$ is a
numerical parameter obtained by integrating over the entire star. The fitting
formula $E_2 = 1.592 \times 10^{-9}\p{M / M_{\odot}}^{2.84}$ as given by
\citet{hurley2002evolution} is commonly used, which varies by many orders of
magnitude for different stars. Moreover, $T_{\rm circ}^{\rm (Zahn)}$ depends on
$M$ and $R$, properties of the entire star, when the tidal torque is entirely
generated at the RCB\@. For these reasons, the expression by \citet{kushnir} is
preferred.

\subsection{Eccentric Binaries}\label{ss:2_eccentric}

The gravitational potential of an eccentric companion at the quadrupole order
can be decomposed as a sum over circular orbits \citep[e.g.][]{sl,vlf}:
\begin{align}
    U\p{\bm{r}, t} &= \sum\limits_{m=-2}^2 U_{2m} \p{\bm{r}, t}
        ,\label{eq:u_ecc}\\
    U_{2m}\p{\bm{r}, t} &= -\frac{GM_2 W_{2m} r^2}{D(t)^3}
        Y_{2m}(\theta, \phi) e^{-imf\!\!\!\:(t)},\nonumber\\
        &= -\frac{GM_2W_{2m} r^2}{a^3}Y_{2m}\p{\theta, \phi}
            \sum\limits_{N = -\infty}^\infty \!\!F_{Nm}e^{-iN\Omega t}
            \label{eq:hansen_decomp}.
\end{align}
Here, the coordinate system is centered on the MS star, $(r, \theta, \phi)$ are
the radial, polar, and azimuthal coordinates of $\bm{r}$ respectively, $W_{2
\pm 2} = \sqrt{3\pi/10}$, $W_{2 \pm 1} = 0$, $W_{20} = -\sqrt{\pi / 5}$, $D(t)$
is the instantaneous distance to the companion, $f$ is the true anomaly,
and $Y_{lm}$ denote the spherical harmonics. $F_{Nm}$ denote the \emph{Hansen
coefficients} for $l = 2$ \citep[also denoted $X^N_{2m}$ in][]{murray1999solar},
which are the Fourier coefficients of the perturbing function, i.e.
\begin{equation}
    \frac{a^3}{D(t)^3} e^{-imf\!\!\!\:(t)} = \sum\limits_{N = -\infty}^\infty
        \!\!F_{Nm} e^{-iN\Omega t}.\label{eq:hansen_series}
\end{equation}
The $F_{Nm}$ can be written explicitly as an integral over the eccentric anomaly
\citep{murray1999solar, sl}:
\begin{equation}
    F_{Nm} = \frac{1}{\pi}\int\limits_{0}^{\pi}
        \frac{\cos\s{N\p{E - e\sin E} - mf(E)}}
            {\p{1 - e\cos E}^2}\;\mathrm{d}E.\label{eq:hansen_integral}
\end{equation}

By considering the effect of each summand in Eq.~\eqref{eq:u_ecc}, the total
torque on the star, energy transfer in the inertial frame, and energy transfer
in the star's corotating frame (which is also the tidal heating rate) can be
obtained \citep{sl, vlf}:
\begin{align}
    T ={}& \sum\limits_{N = -\infty}^\infty F_{N2}^2
        T_{\rm circ}\p{N\Omega - 2\Omega_{\rm s}},\label{eq:tau_sum}
        \\
    \dot{E}_{\rm in} ={}&
        \frac{1}{2}\sum\limits_{N = -\infty}^\infty\Bigg\{
            \p{\frac{W_{20}}{W_{22}}}^2 N\Omega F_{N0}^2 T_{\rm circ}\p{N\Omega}
                \nonumber\\
            &+ N\Omega F_{N2}^2 T_{\rm circ}\p{N\Omega - 2\Omega_{\rm s}}
            \Bigg\},\label{eq:edot_in}\\
    \dot{E}_{\rm rot} ={}& \dot{E}_{\rm in} - \Omega_{\rm s} T
        \label{eq:edot_rot}.
\end{align}
Here, dots indicate time derivatives.

Equations~(\ref{eq:tau_sum}--\ref{eq:edot_in}) can be used to express the
binary orbital decay and circularization rates using
\begin{align}
    \frac{\dot{a}}{a} &= -\frac{2a\dot{E}_{\rm in}}{GMM_2},
        \label{eq:dota}\\
    \frac{\dot{e}e}{1 - e^2} &= -\frac{a\dot{E}_{\rm in}}{GMM_2} +
        \frac{T}{L_{\rm orb}},\label{eq:dote}
\end{align}
where $L_{\rm orb} = MM_2 \s{Ga(1 - e^2) / (M + M_2)}^{1/2}$ is the orbital
angular momentum. The stellar spin synchronization rate can also be computed
assuming that the star rotates rigidly:
\begin{equation}
    \dot{\Omega}_{\rm s}
        = \frac{T}{kMR^2},\label{eq:dots}
\end{equation}
where $kMR^2$ is the moment of inertia of the MS star.

\section{Analytic Evaluation of Tidal Torque and Energy Transfer
Rates}\label{ss:objective}

We can combine the expressions given in Sections~\ref{ss:2_circ}
and~\ref{ss:2_eccentric} to compute the torque and energy transfer rate due to
dynamical tides in an eccentric binary. The tidal torque is obtained by
evaluating Eq.~\eqref{eq:tau_sum} with the circular torque set to
Eq.~\eqref{eq:kushnir_torque}, giving:
\begin{equation}
    T = \sum_{N = -\infty}^{N = \infty} F_{N2}^2 T_0
        \,\mathrm{sgn}\left(N - \frac{2\Omega_{\rm s}}{ \Omega}\right) \left|N -
        \frac{2 \Omega_{\rm s}}{\Omega}\right|^{8/3}.\label{eq:tau_explicit_sum}
\end{equation}
The energy transfer rate in the inertial frame is obtained by evaluating
Eq.~\eqref{eq:edot_in} in the same way, giving:
\begin{align}
    \dot{E}_{\rm in} ={}& \frac{T_0}{2}
        \sum_{N = -\infty}^{\infty} \Bigg[
            N\Omega F_{N2}^2 \mathrm{sgn}\left(N - 2\Omega_{\rm s} / \Omega\right)
                    \left|N - \frac{2 \Omega_{\rm s}}{\Omega}
                    \right|^{8/3}\nonumber\\
            &+ \left(\frac{W_{20}}{W_{22}}\right)^2 \Omega
                    F_{N0}^2 |N|^{11/3}
            \Bigg].\label{eq:ein_explicit_sum}
\end{align}
These two expressions can be used to obtain the orbital decay, circularization, and
spin synchronization rates using Eqs.~(\ref{eq:dota}--\ref{eq:dots}).

While exact, the two sums in
Eqs.~(\ref{eq:tau_explicit_sum}--\ref{eq:ein_explicit_sum}) are difficult to
evaluate for larger eccentricities, where one often must sum hundreds or
thousands of terms, each of which has a different $F_{Nm}$. In the following,
we obtain closed-form approximations to
Eqs.~(\ref{eq:tau_explicit_sum}--\ref{eq:ein_explicit_sum}) when the
eccentricity is large.

\subsection{Approximating Hansen Coefficients}\label{ss:hansens}

To simplify Eqs.~(\ref{eq:tau_explicit_sum}--\ref{eq:ein_explicit_sum}), we seek
tractable approximations for both $F_{N2}$ and $F_{N0}$. Note that while the
Hansen coefficients can be evaluated using the integral expression
Eq.~\eqref{eq:hansen_integral}, this requires calculating a separate integral
for each $N$. Instead, it is more convenient to use the discrete Fourier
Transform of the left hand side of Eq.~\eqref{eq:hansen_series} to calculate
arbitrarily many $N$ at once \citep[as pointed out
by][]{correia2014deformation}. Since $F_{(-N)m} = F_{N(-m)}$, we will only study
the Hansen coefficient behavior for $m \geq 0$.

\subsubsection{$m=2$ Hansen Coefficients}

\begin{figure}
    \centering
    \includegraphics[width=\colummwidth]{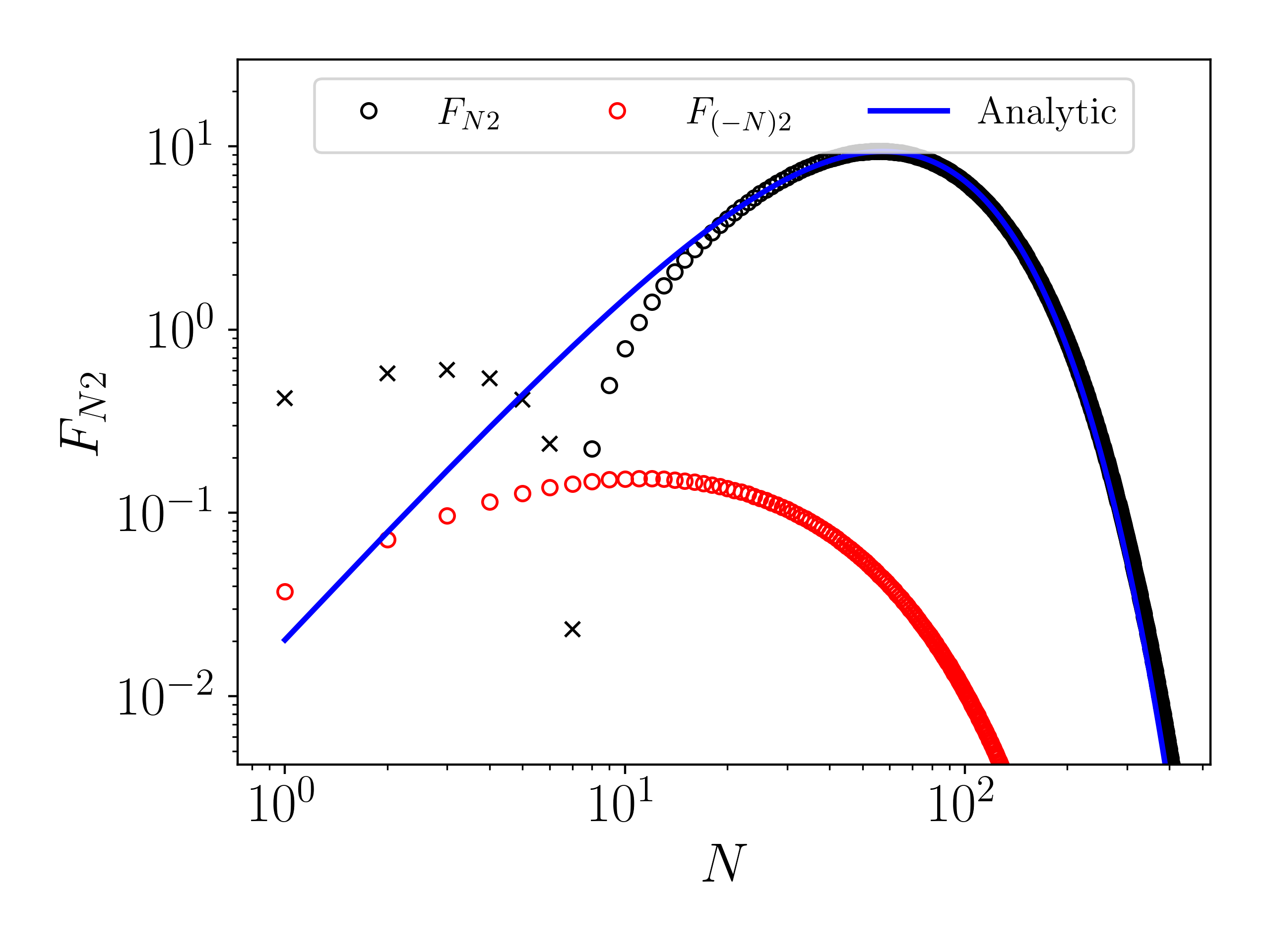}
    \caption{Plot of Hansen coefficients $F_{N2}$ for $e = 0.9$. The red circles
    denote negative $N$, while the black circles and crosses denote positive and
    negative $F_{N2}$. The blue line is the formula given by
    Eq.~\eqref{eq:fn2_fit} with $\eta_2$ and $C_2$ given by
    Eqs.~(\ref{eq:eta2}--\ref{eq:C2}). }\label{fig:hansens}
\end{figure}

Figure~\ref{fig:hansens} shows $F_{N2}$ vs.\ $N$ when $e = 0.9$. First, we note
that $F_{N2}$ is much larger when $N \geq 0$ than for $N < 0$, so we focus on
the behavior for $N \geq 0$. Here, $F_{N2}$ has only one substantial peak. There
are only two characteristic frequency scales: $\Omega$ and $\Omega_{\rm p}$, the
pericentre frequency, defined by
\begin{equation}
    \Omega_{\rm p} \equiv \Omega \frac{\sqrt{1 + e}}{\p{1 - e}^{3/2}}.
        \label{eq:Wperi}
\end{equation}
For convenience, we also define $N_{\rm p}$ as the floor of $\Omega_{\rm p} /
\Omega$, i.e.
\begin{equation}
    N_{\rm p} \equiv \lfloor \Omega_{\rm p} / \Omega\rfloor,
\end{equation}
We find that the peak of the $F_{N2}$ occurs at $N \sim N_{\rm p}$, the only
characteristic scale in $N$ over which $F_{N2}$ can vary. When $N \gg N_{\rm
p}$, the Fourier coefficients must fall off exponentially by the Paley-Wiener
theorem, as the left hand side of Eq.~\eqref{eq:hansen_series} is smooth
\citep[e.g.][]{stein2009real}. When instead $N \ll N_{\rm p}$, there are no
characteristic frequencies between $\Omega$ and $\Omega_{\rm p}$, so we expect
the Hansen coefficients to be scale-free between $N = 1$ and $N_{\rm p}$, i.e.\
a power law in $N$. The expected behaviors in both of these regimes are in
agreement with Fig.~\ref{fig:hansens}.

Motivated by these considerations, we approximate the Hansen coefficients by
\begin{equation}
    F_{N2} \approx
    \begin{cases}
        C_2 N^{p}e^{-N/\eta_2} & N \geq 0,\\
        0 & N < 0,
    \end{cases}\label{eq:fn2_fit}
\end{equation}
for some fitting coefficients $C_2$, $p$, and $\eta_2$. By performing fits to
$F_{N2}$, we find that $p \approx 2$ for substantial
eccentricities, and we fix $p = 2$ for the remainder of this
work\footnote{There is good reason to expect $p = 2$ for $N \ll N_{\rm p}$
as long as the eccentricity is sufficiently large, as then the left-hand side of
Eq.~\eqref{eq:hansen_series} resembles the second derivative of a Dirac delta
function within each orbital period: It is both sharply peaked about $t = 0$ and
has zero derivative three times every period (at $t = \epsilon$, $t = P / 2$,
and $t = P - \epsilon$ for some small $\epsilon \sim \Omega_{\rm p}^{-1}$).
These two characteristics describe the second derivative of a Gaussian with
width $\sim \Omega_{\rm p}^{-1}$. Then, for timescales longer than its width, a
Gaussian resembles a Dirac delta function, which has a flat Fourier spectrum
($\propto N^0$). Finally, since time differentiation multiplies by $N$ in
frequency space, the second derivative of a Gaussian has a Fourier spectrum
$\propto N^2$ for sufficiently small $N \lesssim N_{\rm p}$. As $F_{N2}$ is the
$N$th Fourier coefficient for a function resembling the second derivative of a
Dirac delta function for $N \lesssim N_{\rm p}$, we do indeed expect $F_{N2}
\propto N^2$ in this regime.}.

To constrain the remaining two free parameters $\eta_2$ and $C_2$, we use the
well known Hansen coefficient moments
\begin{align}
    \sum\limits_{N = -\infty}^\infty F_{N2}^2 &= \frac{f_5}{\p{1 - e^2}^{9/2}},
        \label{eq:sum_one}
        \\
    f_5 &\equiv 1 + 3e^2 + \frac{3e^4}{8},\\
    \sum\limits_{N = -\infty}^\infty F_{N2}^2N
        &= \frac{2f_2}{\p{1 - e^2}^6},\label{eq:sum_two}\\
    f_2 &\equiv 1 + \frac{15e^2}{2}
            + \frac{45 e^4}{8} + \frac{5e^6}{16}.
\end{align}
Applying Eq.~\eqref{eq:fn2_fit} to Eqs.~(\ref{eq:sum_one},~\ref{eq:sum_two}), we
obtain the coefficients $\eta_2$ and $C_2$
\begin{align}
    \eta_2 &= \frac{4f_2}{5f_5\p{1 - e^2}^{3/2}}
        \sim \frac{1}{2\p{1 - e}^{3/2}},\label{eq:eta2}\\
    C_2 &= \s{\frac{4f_5}{3\p{1 - e^2}^{9/2}\eta_2^5}}^{1/2}.\label{eq:C2}
\end{align}
Figure~\ref{fig:hansens} illustrates the agreement of Eq.~\eqref{eq:fn2_fit}
using these two values of $\eta_2$ and $C_2$ with the numerical $F_{N2}$. The
good agreement for $N \gtrsim N_{\rm p} / 10$ (which is where $F_{N2}$ is large)
is especially impressive as there are no fitting parameters in
Eq.~\eqref{eq:fn2_fit}, as $C_2$, $\eta_2$, and $p$ are all analytically
constrained. Finally, note that the maximum of the $F_{N2}$ occurs at $N =
\lfloor 2\eta_2 \rfloor \sim \p{1 - e}^{-3/2} \sim N_{\rm p}$.

\subsubsection{$m = 0$ Hansen Coefficients}

\begin{figure}
    \centering
    \includegraphics[width=0.9\colummwidth]{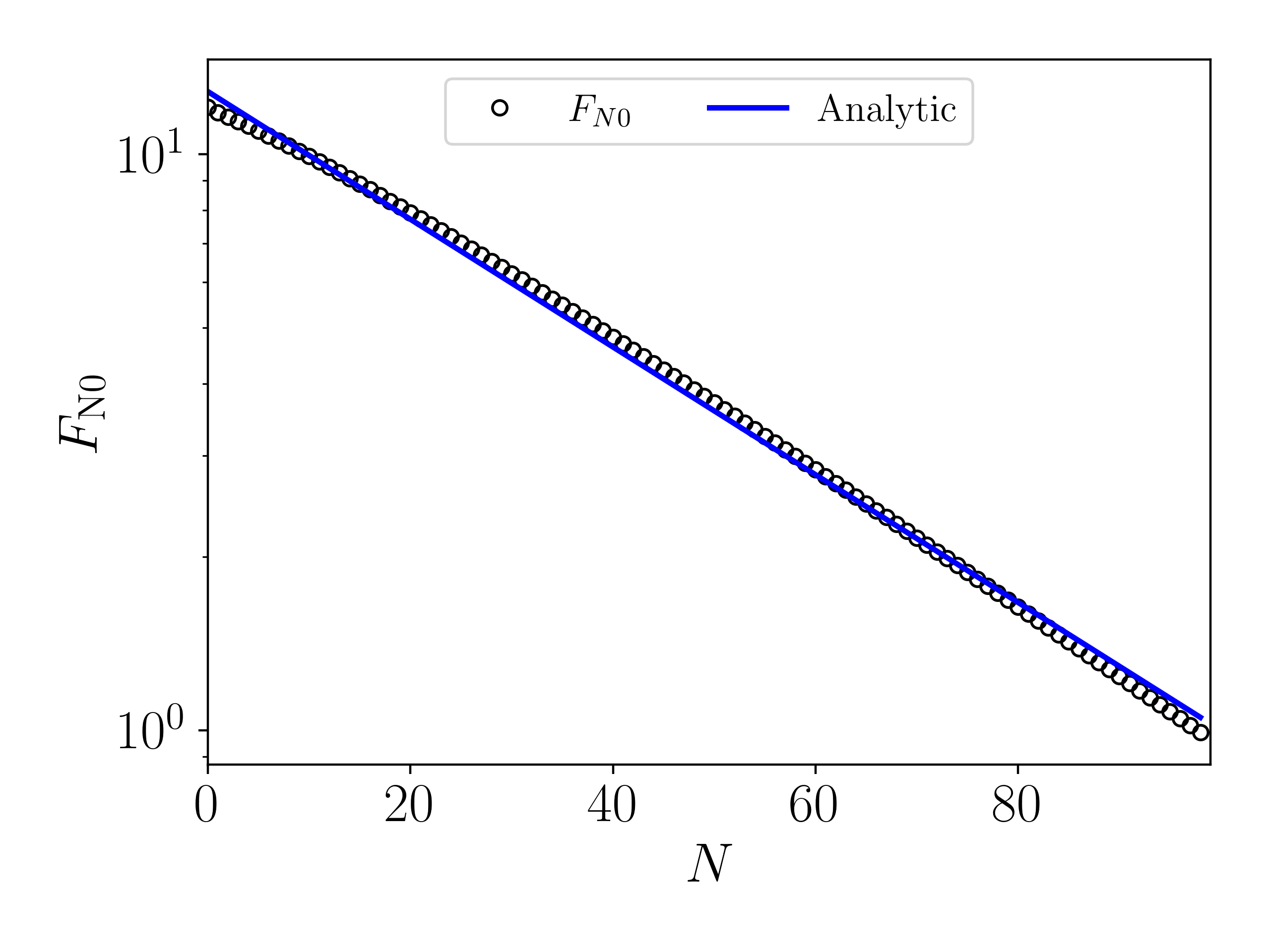}
    \caption{Plot of $F_{N0}$ (black circles) for $e = 0.9$. Since $F_{N0} =
    F_{(-N)0}$, we only show positive $N$. The blue line is given by
    Eq.~\eqref{eq:fn0_fit} with $\eta_0$ and $C_0$ given by
    Eqs.~(\ref{eq:eta0}--\ref{eq:C0}).}\label{fig:fn0_fit}
\end{figure}

We now turn to the $m = 0$ Hansen coefficients, $F_{N0}$, which are shown in
Fig.~\ref{fig:fn0_fit}. We know that $F_{N0} = F_{(-N)0}$, so we consider only
$N \geq 0$. From the figure, we see that the $F_{N0}$ decay exponentially. There
is only one characteristic scale available for this decay, namely $N_{\rm p}$.
Therefore, we naturally assume the $F_{N0}$ coefficients can be approximated by
a function of form:
\begin{equation}
    F_{N0} = C_0 e^{-\abs{N} / \eta_0}.\label{eq:fn0_fit}
\end{equation}
The two free parameters $C_0$ and $\eta_0$ are constrained by the identities
\begin{align}
    \sum\limits_{N = -\infty}^\infty F_{N0}^2 &= \frac{f_5}{\p{1 - e^2}^{9/2}}
        \label{eq:e_sum1}
        ,\\
    \sum\limits_{N = -\infty}^\infty F_{N0}^2 N^2
        &= \frac{9e^2}{2\p{1 - e^2}^{15/2}}
            f_3,\label{eq:e_sum2}\\
    f_3 &= \frac{1}{2} + \frac{15e^2}{8} + \frac{15 e^4}{16}
        + \frac{5e^6}{128}.
\end{align}
Applying Eq.~\eqref{eq:fn0_fit} to Eqs.~(\ref{eq:e_sum1}--\ref{eq:e_sum2}), we
obtain
\begin{align}
    \eta_0 &= \s{\frac{9e^2f_3}{\p{1 - e^2}^{3}f_5}}^{1/2},\label{eq:eta0}\\
    C_0 &= \s{\frac{f_5}{\p{1 - e^2}^{9/2}\eta_0}}^{1/2}.\label{eq:C0}
\end{align}
Figure~\ref{fig:fn0_fit} illustrates the agreement of Eq.~\eqref{eq:fn0_fit}
using these two values of $\eta_0$ and $C_0$with the numerically computed
$F_{N0}$.

\subsection{Approximate Expressions for Torque and Energy Transfer}\label{s:eval}

Having found good approximations for the Hansen coefficients, we now apply them
to simplify the expressions for the torque and the energy transfer rate in
Eqs.~(\ref{eq:tau_explicit_sum}--\ref{eq:ein_explicit_sum}).

\subsubsection{Tidal Torque}\label{ss:torque_eval}

To simplify Eq.~\eqref{eq:tau_explicit_sum}, we replace $F_{N2}$ with
Eq.~\eqref{eq:fn2_fit} and the sum with an integral, obtaining
\begin{equation}
    T \approx T_0 \int_0^\infty C_2^2 N^4 e^{-2N / \eta_2}
        \sgn\left(N - 2\Omega_{\rm s} / \Omega\right) \left|N - 2 \Omega_{\rm s} /
            \Omega\right|^{8/3}\;\mathrm{d}N.\label{eq:tau_int}
\end{equation}

This expression is already easier to evaluate than
Eq.~\eqref{eq:tau_explicit_sum}, but we can use further approximations to obtain
a closed form. We first analyze Eq.~\eqref{eq:tau_int} in the small-spin limit,
where it can be integrated analytically\footnote{The key to the success of our
approach is that sums of form $\sum_{n = -\infty}^\infty F_{N2}^2 N^p$ can be
approximated for non-integer $p$ in terms of the $\Gamma$ function, since
$\int_0^\infty x^pe^{-x}\;\mathrm{d}x = \Gamma(p - 1)$.}, giving
\begin{align}
    T\p{\abs{\Omega_{\rm s}} \ll \Omega_{\rm p}} &\simeq T_0 \frac{f_5(e)
        (\eta_2/2)^{8/3}}{(1 - e^2)^{9/2}} \frac{\Gamma(23/3)}{4!}.
        \label{eq:tau_lowspin}
\end{align}
Note that this has the scaling $T \sim T_0\p{1 - e}^{-17/2} \sim T_{\rm p}
\Omega / \Omega_{\rm p}$, where $T_{\rm p}$ is the torque exerted by a circular
orbit with separation equal to the pericentre separation $a_{\rm p} \equiv a(1 -
e)$, i.e.
\begin{equation}
    T_{\rm p} = \beta_2\frac{GM_2^2r_{\rm c}^5}{a_{\rm p}^6}
            \p{\frac{\Omega_{\rm p}}{\sqrt{GM_{\rm c}/r_{\rm c}^3}}}^{8/3}
            \frac{\rho_{\rm c}}{\bar{\rho}_{\rm c}} \p{1 - \frac{\rho_{\rm
            c}}{\bar{\rho}_{\rm c}}}^2
            \sim T_0\p{1 - e}^{-10}.\label{eq:def_tperi}
\end{equation}

\begin{figure}
    \centering
    \includegraphics[width=\colummwidth]{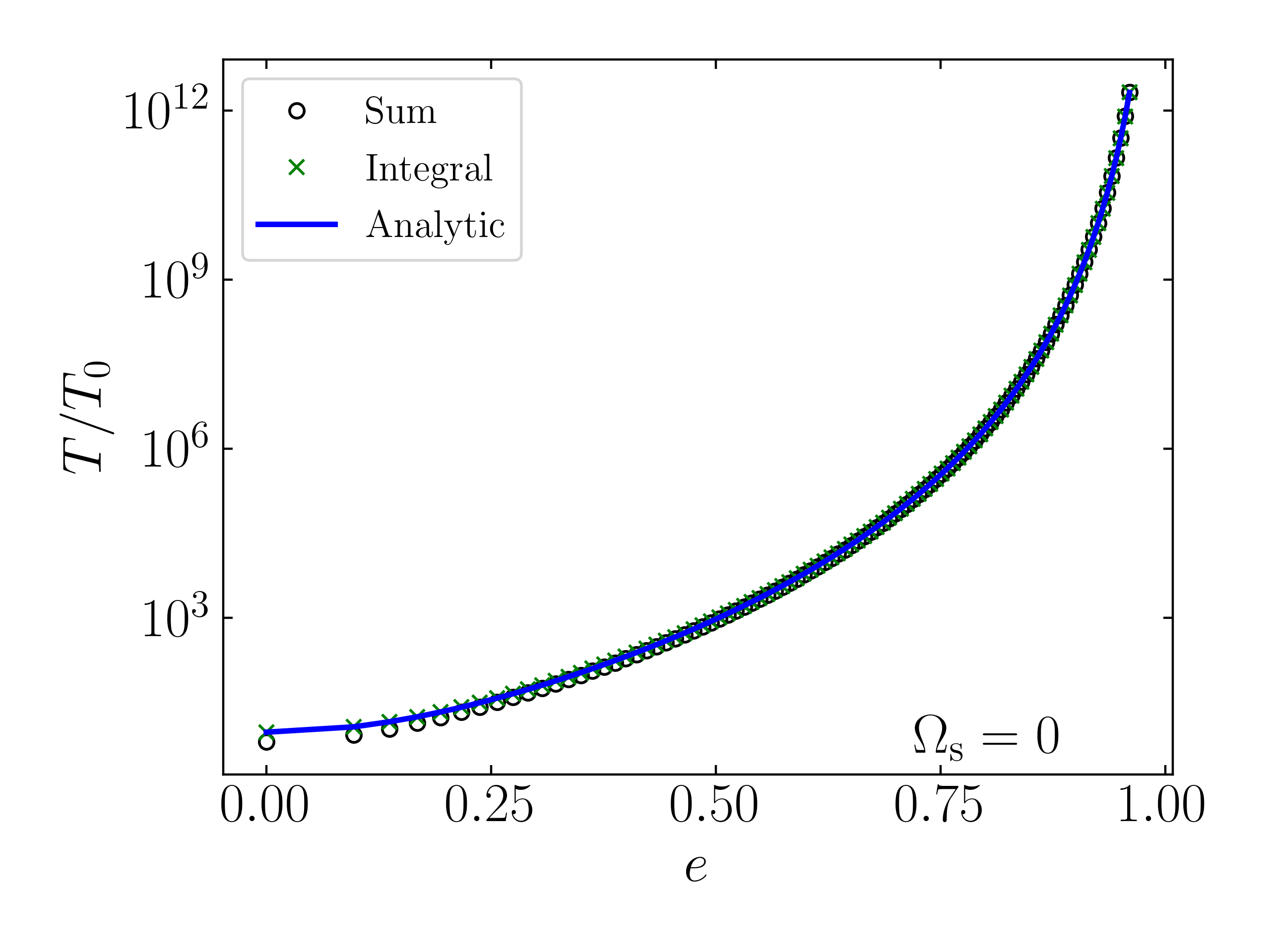}
    \includegraphics[width=\colummwidth]{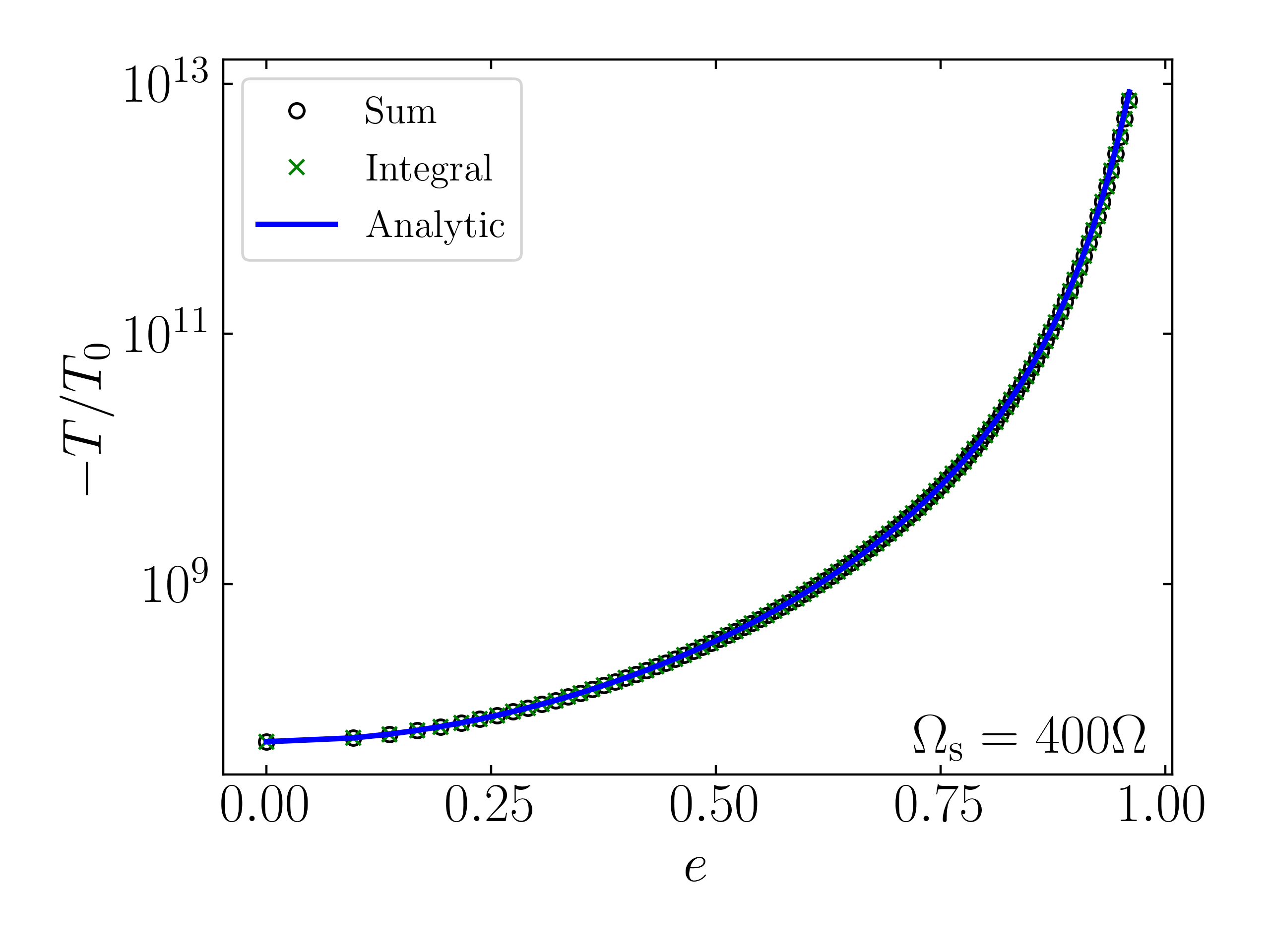}
    \caption{The tidal torque on a non-rotating (top) and rapidly rotating (bottom)
    star due to a companion with orbital eccentricity $e$. Black circles
    represent direct summation of Eq.~\eqref{eq:tau_explicit_sum}, green
    crosses are evaluated using the integral approximation
    Eq.~\eqref{eq:tau_int}, and the blue line is Eq.~\eqref{eq:tau_approx}. In
    the small and large spin limits, Eq.~\eqref{eq:tau_approx} reduces to
    Eq.~\eqref{eq:tau_lowspin} and Eq.~\eqref{eq:tau_highspin} respectively.
    }\label{fig:totals_ecc0}
\end{figure}

The top panel of Fig.~\ref{fig:totals_ecc0} compares Eq.~\ref{eq:tau_lowspin} to
the integral of Eq.~\eqref{eq:tau_int} and to the direct sum of
Eq.~\eqref{eq:tau_explicit_sum} as a function of the eccentricity. It can be see
that both the integral and the analytic closed form perform well for
moderate-to-large eccentricities, but both over-predict the torque at small $e
\lesssim 0.3$. This discrepancy is expected: there are only a few non-negligible
summands in Eq.~\eqref{eq:tau_explicit_sum} when $e$ is small, so replacing the
sum over $N$ with an integral is expected to introduce significant inaccuracy
that appears in both the integral and closed-form expressions.

Eq.~\eqref{eq:tau_lowspin} is valid so long as $\abs{\Omega_{\rm s} / \Omega} \ll
N_{\max}$ (or equivalently, $\abs{\Omega_{\rm s} \ll \Omega_{\rm p}}$), where
$N_{\max} = 10 \eta_2/3$ is where the integrand is in Eq.~\eqref{eq:tau_int}
maximized. If instead $\abs{\Omega_{\rm s} / \Omega} \gg N_{\max}$, the torque
can be evaluated directly using Eq.~\eqref{eq:tau_explicit_sum} and the known
Hansen coefficient moments, giving:
\begin{equation}
    T\p{\abs{\Omega_{\rm s}} \gg \Omega_{\rm p}}
        \simeq -T_0 \sgn (\Omega_{\rm s})\;\left|2
        \Omega_{\rm s} / \Omega\right|^{8/3} \frac{f_5(e)}{(1 -
        e^2)^{9/2}}.\label{eq:tau_highspin}
\end{equation}
The bottom panel of Fig.~\ref{fig:totals_ecc0} compares this formula to the
integral of Eq.~\eqref{eq:tau_int} and to the direct sum of
Eq.~\eqref{eq:tau_explicit_sum} as a function of the eccentricity, where
$\Omega_{\rm s} / \Omega = 400$. Here, $N_{\max} \ll 400$ for all eccentricities
shown. We see that direct summation, the integral expression, and
Eq.~\eqref{eq:tau_highspin} agree very well for all eccentricities.

\begin{figure}
    \centering
    \includegraphics[width=\colummwidth]{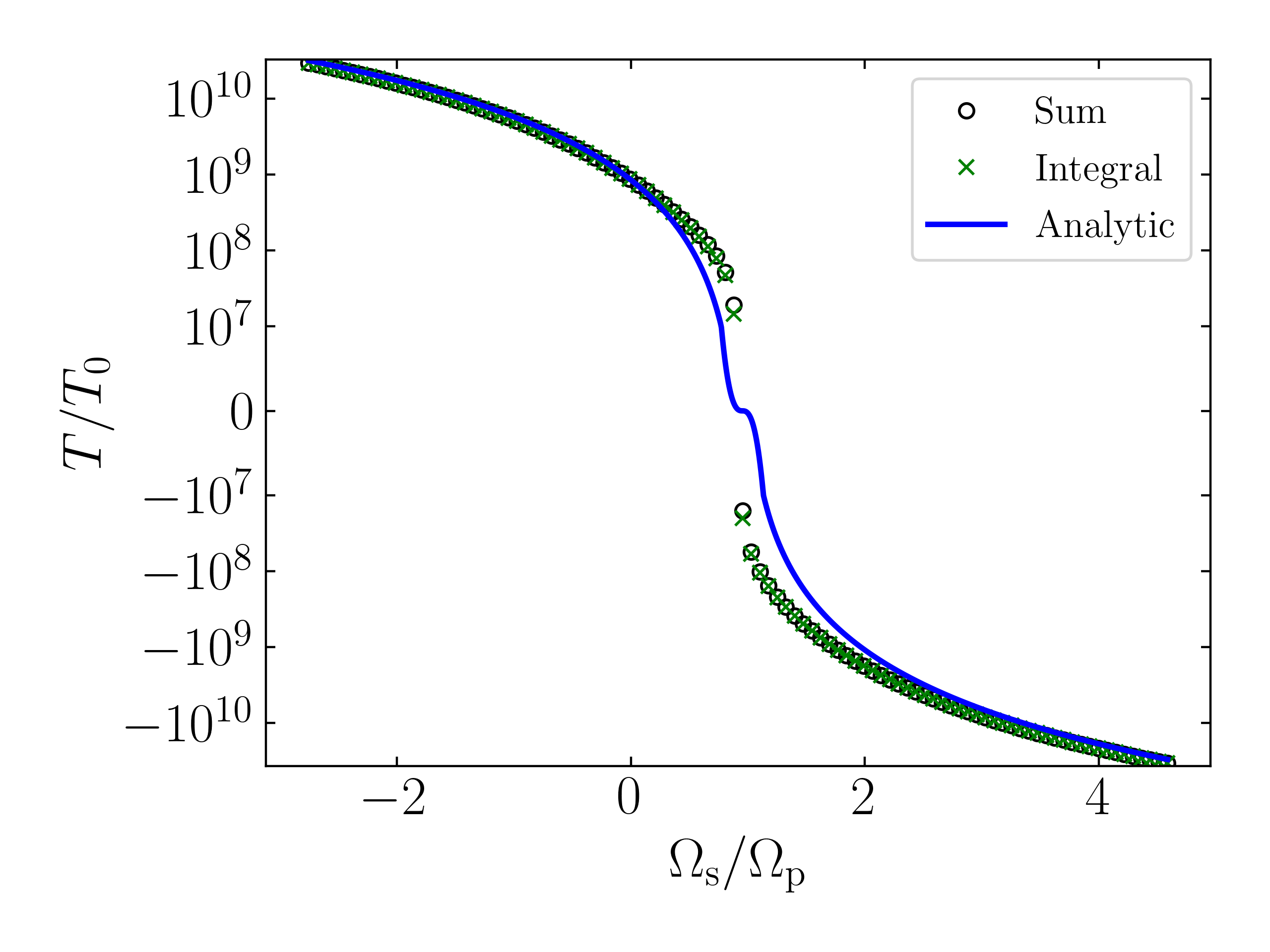}
    \caption{The tidal torque as a function of the stellar spin for a highly
    eccentric $e = 0.9$ companion. The black circles represent direct summation
    of Eq.~\eqref{eq:tau_explicit_sum}, green crosses the integral approximation
    (Eq.~\ref{eq:tau_int}), and the solid line the analytic closed-form
    expression Eq.~\eqref{eq:tau_approx}. The spin is normalized by the
    pericentre orbital frequency $\Omega_{\rm p} \approx 43 \Omega$
    (Eq.~\ref{eq:Wperi}). }\label{fig:totals_s}
\end{figure}

Having obtained closed-form expressions of Eq.~\eqref{eq:tau_int} for small
and large spins, we can further derive a single expression joining these two
limits. To do this, we first assume that the spin is small but non-negligible.
In this regime, we make the approximation
\begin{equation}
    N - 2\Omega_{\rm s} / \Omega \simeq \frac{N}{N_{\max}}
        \left(N_{\max} - \frac{2\gamma_T
        \Omega_{\rm s}}{\Omega}\right)\label{eq:nmax_ansatz},
\end{equation}
for some free parameter $\gamma_T$. With this, we can integrate
Eq.~\eqref{eq:tau_int} in closed form. We can fix $\gamma_T$ by requiring our
expression reproduce the large spin limit (Eq.~\ref{eq:tau_highspin}) when
taking $\abs{\Omega_{\rm s}} \to \infty$. This procedure gives an expression for
the torque that agrees with both limiting forms
(Eq.~\ref{eq:tau_lowspin}--\ref{eq:tau_highspin}) and is given by
\begin{align}
    T &\simeq T_0 \frac{f_5(e) (\eta_2/2)^{8/3}}{(1 - e^2)^{9/2}}
        \;\mathrm{sgn}\left(1 - \gamma_T\frac{\Omega_{\rm s}}{\eta_2\Omega}\right)
            \left|
                \frac{4}{\gamma_T}
                \p{1 - \gamma_T\frac{\Omega_{\rm s}}{\eta_2\Omega}}
            \right|^{8/3} ,\label{eq:tau_approx}
\end{align}
where
\begin{align}
    \gamma_T &= 4\p{\frac{4!}{\Gamma(23/3)}}^{3/8} \approx 0.691.
\end{align}
Figure~\ref{fig:totals_s} compares this expression to the integral of
Eq.~\eqref{eq:tau_int} and to the direct sum of Eq.~\eqref{eq:tau_explicit_sum}
at fixed $e = 0.9$ and varying $\Omega_{\rm s}$. We see that
Eq.~\eqref{eq:tau_approx} agrees well with the integral and sum for both large
and small spins, and is also reasonably accurate for intermediate spins.
However, Eq.~\eqref{eq:tau_int} is more accurate than Eq.~\eqref{eq:tau_approx}
when $T$ changes signs and $\abs{T}$ is small. This is also expected: $T$
changes signs when the spin approaches \emph{}
(Section~\ref{ss:}) because large contributions to the sum
in Eq.~\eqref{eq:tau_explicit_sum} have opposite signs and mostly cancel out.
Thus, small inaccuracies in the summand result in significant discrepancies in
the total torque. The integral approximation, Eq.~\eqref{eq:tau_int}, is
expected to be in good agreement with the direct sum,
Eq.~\eqref{eq:tau_explicit_sum}, as the accuracy of the Hansen coefficient
approximation in Section~\ref{ss:hansens} is good for large eccentricities, thus
guaranteeing term-by-term accuracy. On the other hand, the closed-form
expression, Eq.~\eqref{eq:tau_approx}, is more approximate when $\Omega_{\rm s}
/ \Omega \sim N_{\max} \sim \eta_2$. In fact, Eq.~\eqref{eq:tau_approx} predicts
$\rdil{T}{\Omega_{\rm s}} \approx 0$ near . This is not
accurate and is an artifact of our factorization ansatz in
Eq.~\eqref{eq:nmax_ansatz}.

In summary, the tidal torque must be evaluated with explicit summation
(Eq.~\ref{eq:tau_explicit_sum}) when $e\lesssim 0.3$ (see discussion after
Eq.~\ref{eq:tau_lowspin}), can be approximated by the
integral expression (Eq.~\ref{eq:tau_int}) when $e$ is large for all values of
$\Omega_{\rm s}$, and otherwise can be approximated by the closed-form
expression given by Eq.~\eqref{eq:tau_approx}. Recall that when $e$ is small,
the explicit summation of Eq.~\eqref{eq:tau_explicit_sum} is quite simple, as
good accuracy can be obtained with just the first few terms in the summation.

\subsubsection{Pseudosynchronization}\label{ss:}

\begin{figure}
    \centering
    \includegraphics[width=\colummwidth]{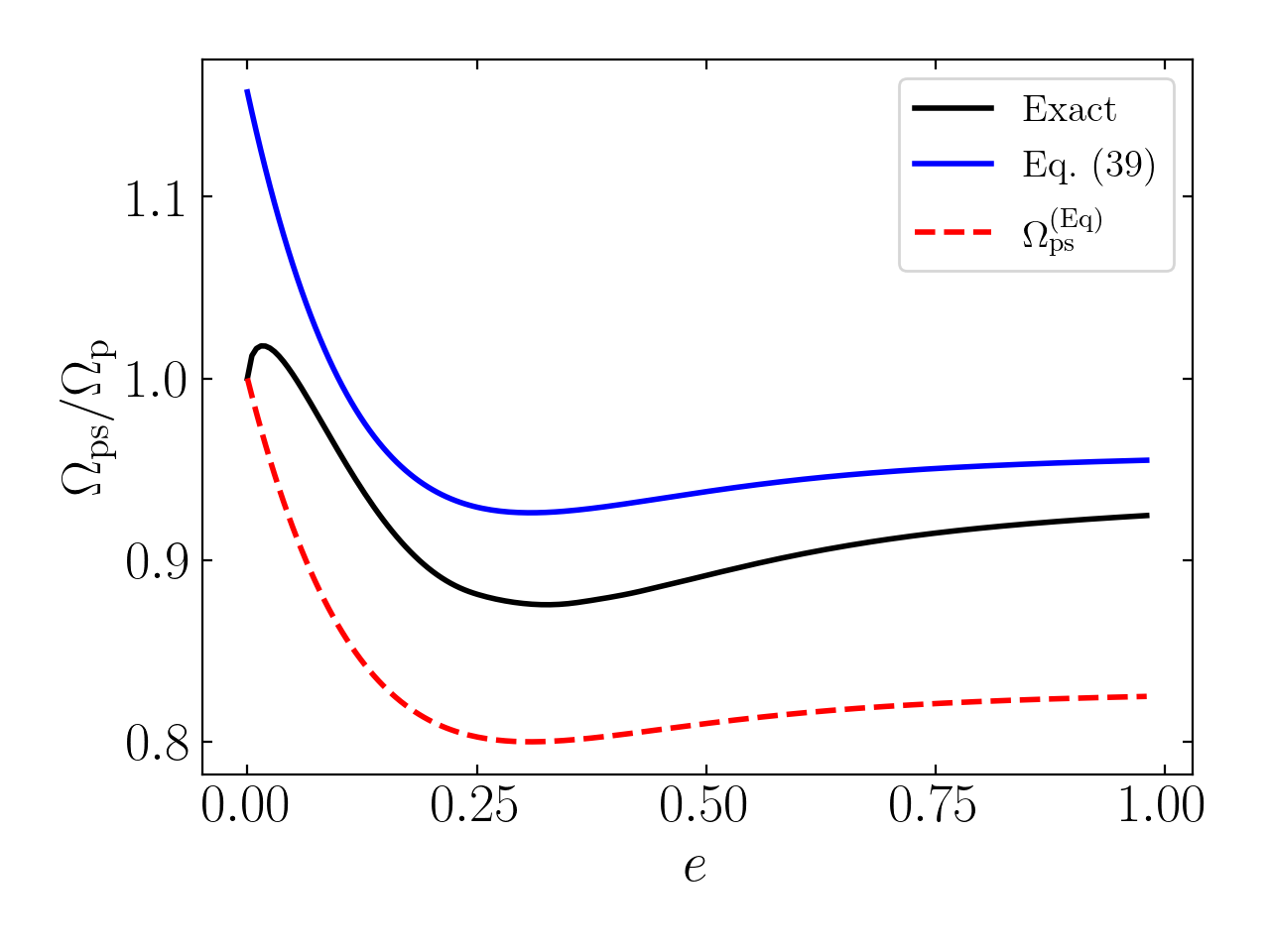}
    \caption{The  spin frequencies $\Omega_{\rm ps}$
    normalized by the pericentre frequency $\Omega_{\rm p}$ (Eq.~\ref{eq:Wperi})
    as a function of eccentricity. The blue line is given by
    Eq.~\eqref{eq:eta2_691}. The black line shows the exact solution, obtained
    by using a root finding algorithm to solve for the zero of
    Eq.~\eqref{eq:tau_sum}. The red dashed line shows the
    spin frequency predicted by the weak friction theory of equilibrium tides
    (Eq.~\ref{eq:weaktide}).
    }\label{fig:pseudosync}
\end{figure}

In general, the exact torque as given by Eq.~\eqref{eq:tau_explicit_sum}
vanishes for a single $\Omega_{\rm s}$, which we call the
\emph{pseudo-synchronized} spin frequency. An approximation for the
pseudo-synchronized spin can be directly read off from Eq.~\eqref{eq:tau_approx}:
\begin{equation}
    \frac{\Omega_{\rm ps}}{\Omega} =
        \frac{\eta_2}{\gamma_T} = \frac{4f_2(e)}{5\gamma_T f_5(e)\p{1 -
        e^2}^{3/2}}.\label{eq:eta2_691}
\end{equation}
This has the expected scaling $\Omega_{\rm ps} \sim \Omega_{\rm p}$ (the
pericentre orbital frequency). Figure~\ref{fig:pseudosync} compares this
prediction for the pseudo-synchronized spin to the exact one obtained by applying
a root finding algorithm to Eq.~\eqref{eq:tau_explicit_sum}. We see that
Eq.~\eqref{eq:eta2_691} is a good approximation for the pseudo-synchronized spin
frequency when $e \gtrsim 0.1$.

In passing, we note that, in the standard weak friction theory of equilibrium
tides, the pseudo-synchronized spin is given by \citep{alexander73, hut81}
\begin{equation}
    \frac{\Omega_{\rm ps}^{\rm (Eq)}}{\Omega} = \frac{f_2(e)}{f_5(e)\p{1 -
        e^2}^{3/2}}.\label{eq:weaktide}
\end{equation}
Though describing a different tidal phenomenon, this only differs from
Eq.~\eqref{eq:eta2_691} by a factor of $4 / (5\gamma_{T}) \approx 1.15$. We show
it for comparison as the red dotted line in Fig.~\ref{fig:pseudosync}.

\subsubsection{Energy Transfer}

We now turn our attention to Eq.~\eqref{eq:edot_in} and replace $F_{N2}$ and
$F_{N0}$ with their respective approximations (Eqs.~\ref{eq:fn2_fit}
and~\ref{eq:fn0_fit}) to obtain the energy transfer rate
\begin{align}
    \dot{E}_{\rm in} ={}&
        \frac{T_0 \Omega}{2} \int\limits_0^\infty \Big[
            C_2^2 N^5 e^{-2N/\eta_2} \mathrm{sgn}\left(N - 2\Omega_{\rm s} /
                \Omega\right) \left|N - 2 \Omega_{\rm s} /
                \Omega\right|^{8/3}\nonumber\\
            &+ 2 \p{\frac{W_{20}}{W_{22}}}^2 C_0^2 e^{-2N / \eta_0} N^{11/3}
        \Big]\;\mathrm{d}N.\label{eq:ein_int}
\end{align}
We evaluate the $m = 2$ and $m = 0$ contributions to this expression separately.

We first examine the $m = 2$ contribution using the same procedure in
Section~\ref{ss:torque_eval} for the torque. If the spin is moderate, i.e.\
$\abs{\Omega_{\rm s} / \Omega} \lesssim N_{\max}$ where now $N_{\max} = 23
\eta_2 / 6$, we make the approximation
\begin{equation}
    N - 2\Omega_{\rm s} / \Omega \simeq \frac{N}{N_{\max}}
        \left(N_{\max} - \frac{2\gamma_E
        \Omega_{\rm s}}{\Omega}\right)\label{eq:nmax_ansatz2},
\end{equation}
where $\gamma_E$ is a free parameter. This lets us integrate the $m = 2$
contribution to Eq.~\eqref{eq:ein_int} analytically. We constrain $\gamma_E$ by
requiring agreement with the large-spin limit: for $\abs{\Omega_{\rm s} /
\Omega} \gg N_{\max}$, we have
\begin{equation}
    \dot{E}_{\rm in}^{(m=2)} \p{\abs{\Omega_{\rm s}} \gg \Omega_{\rm p}}
        \simeq -\frac{T_0\Omega}{2} \; \mathrm{sgn}(\Omega_{\rm s})|2\Omega_{\rm
        s}/\Omega|^{8/3} \frac{2f_2(e)}{(1 - e^2)^6}.\label{eq:ein_highs}
\end{equation}
This fixes $\gamma_E$ and we obtain the complete $m = 2$ contribution to
Eq.~\eqref{eq:ein_int}:
\begin{align}
    \dot{E}_{\rm in}^{(m=2)}
        \simeq{}& \frac{T_0\Omega f_5(e) (\eta_2/2)^{11/3}}{2(1-e^2)^{9/2}}
            \nonumber\\
        &\times \mathrm{sgn}\p{1 - \gamma_E\frac{\Omega_{\rm s}}{\eta_2 \Omega}}
            \left|
                \frac{4}{\gamma_E}
                \p{1 - \gamma_E\frac{\Omega_{\rm s}}{\eta_2 \Omega}}
            \right|^{8/3}\label{eq:ein_2_comp},
\end{align}
where
\begin{align}
    \gamma_E ={}& 4\p{\frac{5!}{\Gamma\p{26/3}}}^{3/8}
        \approx 0.5886.
\end{align}
Note that, when $\Omega_{\rm s} \approx 0$, Eq.~\eqref{eq:ein_2_comp} gives the
expected scaling of $\dot{E}_{\rm in}^{\rm (m=2)} \sim T_{\rm p}\Omega$ where
$T_{\rm p}$ is the torque exerted by a circular companion at the pericentre
separation, given by Eq.~\eqref{eq:def_tperi}.

The $m = 0$ contribution to Eq.~\eqref{eq:ein_int} can be straightforwardly
integrated using the parameterization Eq.~\eqref{eq:fn0_fit}. The sum of the two
contributions then gives the total energy transfer rate:
\begin{align}
    \dot{E}_{\rm in} ={}& \frac{T_0 \Omega}{2}\Bigg[\frac{f_5(e)
        (\eta_2/2)^{11/3}}{(1-e^2)^{9/2}} \nonumber\\
        &\times \mathrm{sgn}\p{1 - \gamma_E\frac{\Omega_{\rm s}}{\eta_2 \Omega}}
            \left|\frac{4}{\gamma_{E}}
            \p{1 - \gamma_E\frac{\Omega_{\rm s}}{\eta_2 \Omega}}\right|^{8/3}
            \nonumber\\
        &+
    \frac{f_5(e) \Gamma(14 / 3)}{(1 - e^2)^{10}} \left(\frac{3}{2}\right)^{8/3}
            \left(\frac{e^2 f_3(e)}{f_5(e)}\right)^{11/6}\Bigg].
            \label{eq:ein_dot_tot}
\end{align}

\begin{figure}
    \centering
    \includegraphics[width=\colummwidth]{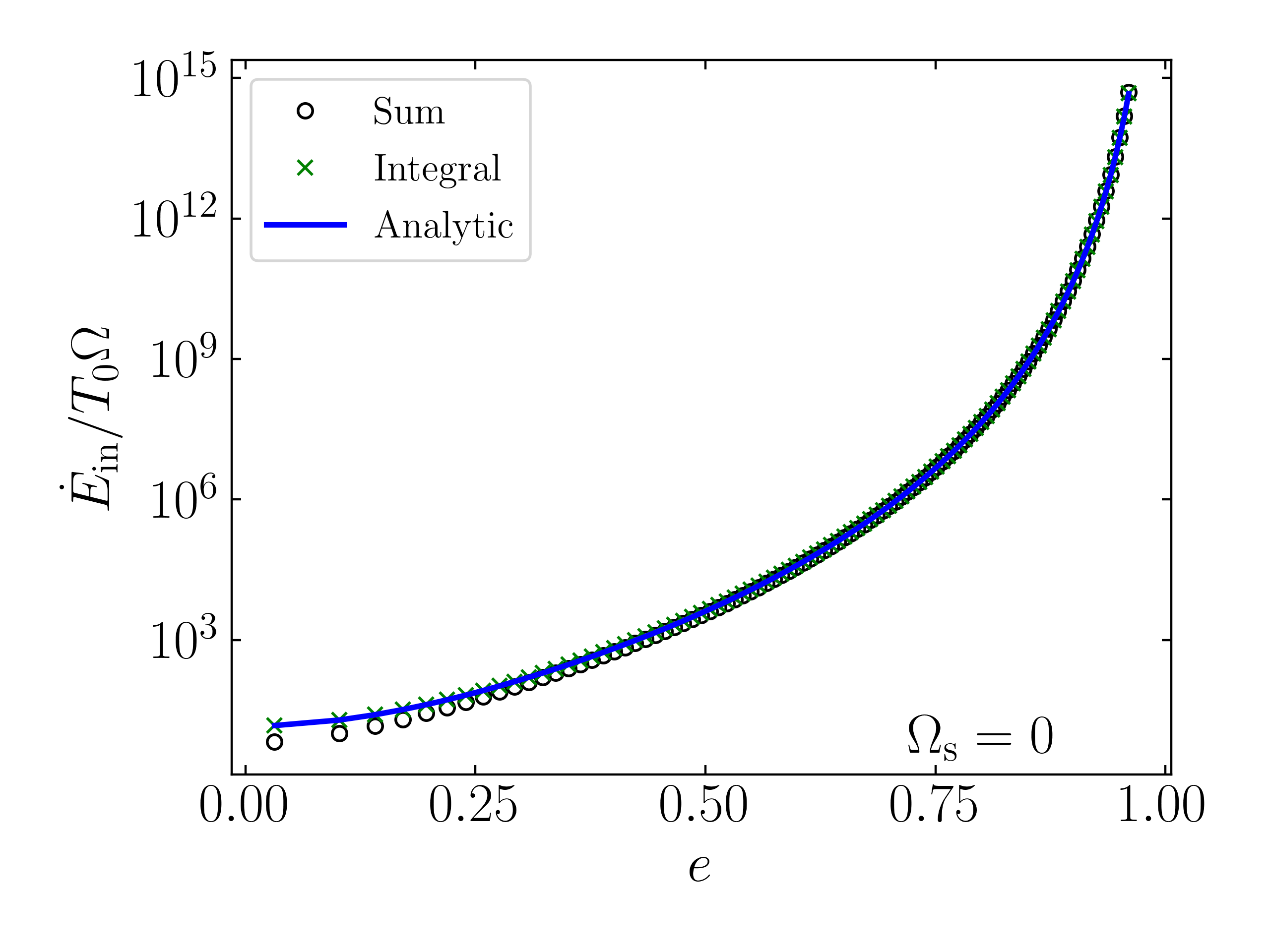}
    \includegraphics[width=\colummwidth]{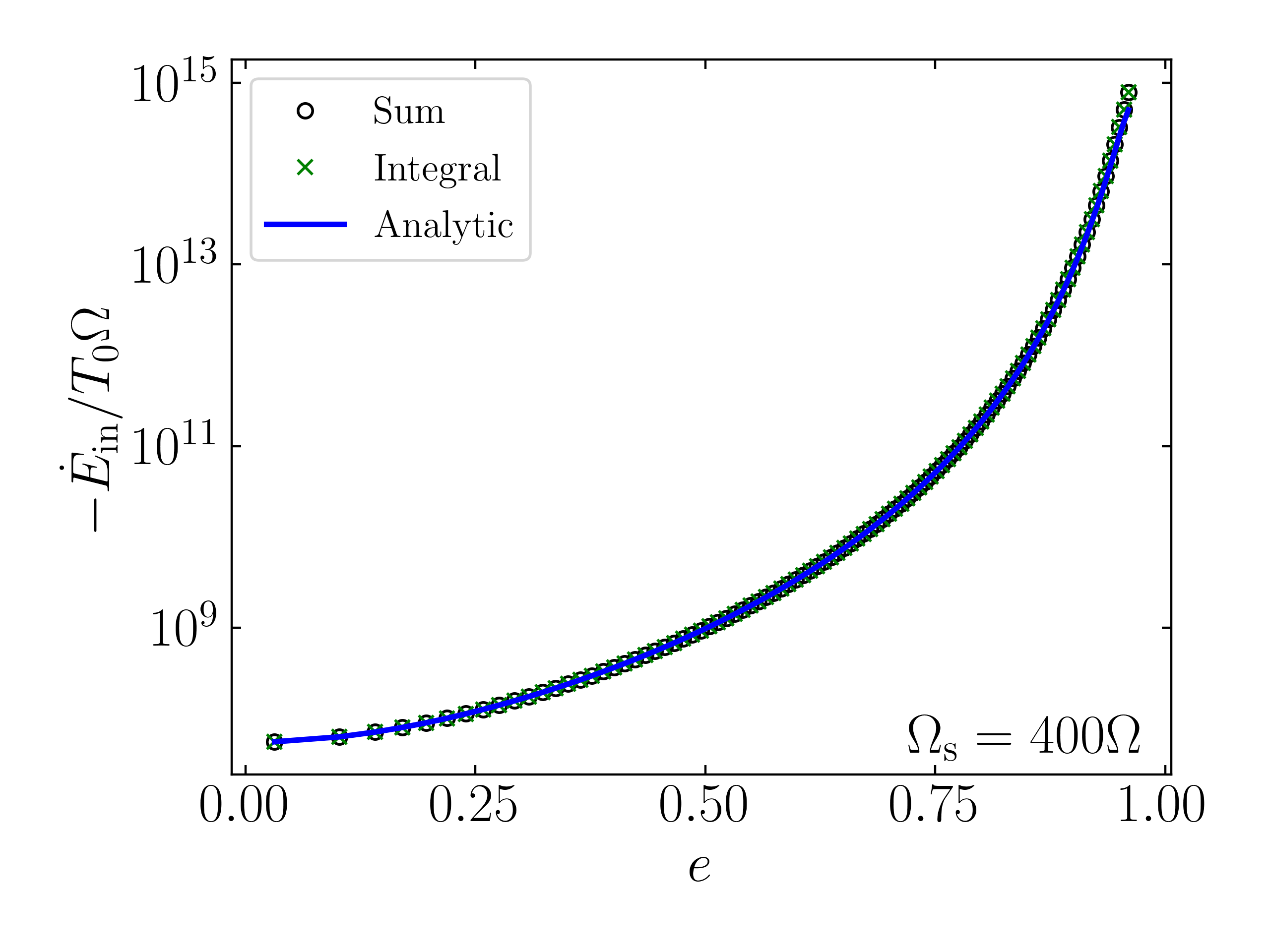}
    \caption{The tidal energy transfer rate $\dot{E}_{\rm in}$ for a
    non-rotating (top) and a rapidly rotating (bottom) star. The black circles
    represent direct summation of Eq.~\eqref{eq:ein_explicit_sum}, green crosses
    the integral form Eq.~\eqref{eq:ein_int}, and the blue line the closed-form
    expression Eq.~\eqref{eq:ein_dot_tot}. }\label{fig:e0}
\end{figure}

\begin{figure}
    \centering
    \includegraphics[width=\colummwidth]{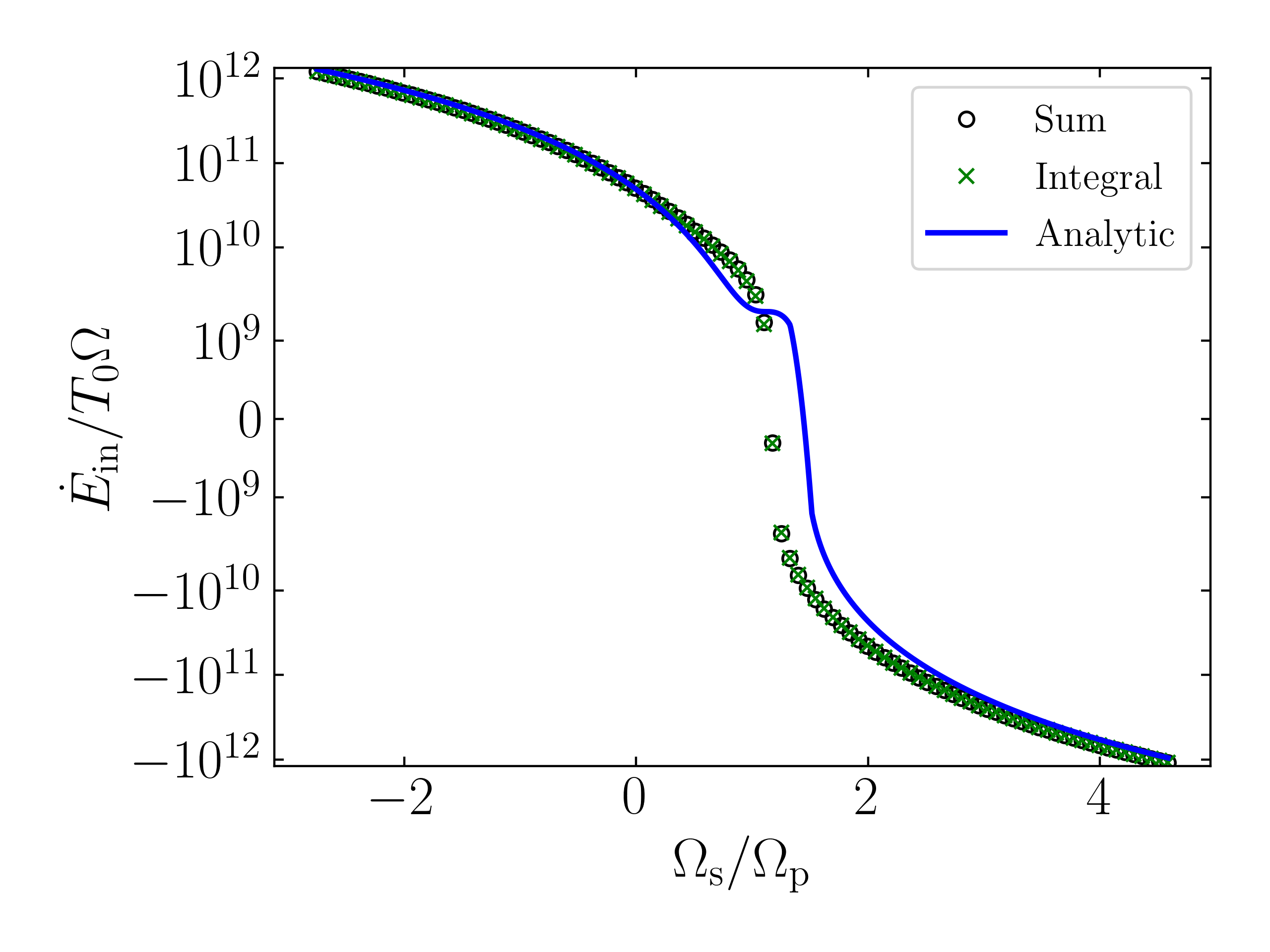}
    \caption{The tidal energy transfer rate $\dot{E}_{\rm in}$ as a function of
    spin [normalized by $\Omega_{\rm p}$; Eq.~\eqref{eq:Wperi}] for a highly
    eccentric $e = 0.9$ companion. The black circles represent direct summation
    of Eq.~\eqref{eq:edot_in}, green crosses the integral approximation
    Eq.~\eqref{eq:ein_int}, and the blue line the analytic closed form
    Eq.~\eqref{eq:ein_dot_tot}. }\label{fig:e_spin}
\end{figure}

The two panels of Fig.~\ref{fig:e0} compare this expression with the integral
form Eq.~\eqref{eq:ein_int} and the direct sum Eq.~\eqref{eq:ein_explicit_sum}
for small and large spins. The agreements are excellent except that, when the
spin and eccentricity are small, both the integral and closed form expressions
over-predict the energy transfer rate. Figure~\ref{fig:e_spin} compares these
three expressions as a function of spin for $e = 0.9$. The performance of
Eq.~\eqref{eq:ein_dot_tot} degrades when the system is near
 ($\Omega_{\rm s} \simeq \Omega_{\rm p}$), but generally
captures the correct behavior of the exact result, while Eq.~\eqref{eq:ein_int}
is accurate for all spins. As was the case with the tidal torque, we see that
the evaluation of the energy transfer rate when $e \lesssim 0.3$ requires direct
summation (Eq.~\ref{eq:ein_explicit_sum}), and the evaluation when $e$ is
substantial but the spin is near  can be performed using
the integral approximation (Eq.~\ref{eq:ein_int}), and otherwise the evaluation
can be performed using the closed-form expression (Eq.~\ref{eq:ein_dot_tot}).

\section{PSR J0045-7319/B-Star Binary}\label{s:j00457319}

As an application of our analytical results above, we consider the PSR
J0045-7319/MS binary system \citep{kaspi1994massive} and attempt to explain its
orbital decay via dynamical tides. This system is one of the few pulsar binaries
discovered so far that have massive MS companions \citep[the other known
binaries are, PSR B1259-63, PSR J1740-3052, PSR J1638-4725, J2032+4127;][]{nsms1,
nsms3, nsms2, nsms4}. These systems evolve from MS-MS binaries when one of the
stars explodes in a supernova to form a neutron star (NS), and will eventually
become double NS systems when the MS companion also explodes to form a NS\@.
Thus, characterizing the evolution of such systems is important for
understanding the formation of double NS binaries \citep[e.g.][]{tauris2017}.
The PSR J0045-7319 system contains a radio pulsar ($M_2\simeq 1.4M_\odot$) and a
massive B star ($M \simeq 8.8M_{\odot}$) companion in an eccentric ($e=0.808$)
orbit with period $P =51.17\;\mathrm{days}$, corresponding to a semi-major axis
$a = 126R_{\odot}$. Timing observation shows that the orbit is decaying at the
rate $\dot{P}= -3.03\times 10^{-7}$, or $P/\dot P\simeq -0.5$~Myr
\citep{kaspi1996params}. The measured B-star luminosity $L = 1.2 \times
10^4L_{\odot}$ and surface temperature $T_{\rm surf} = (24000 \pm
1000)\;\mathrm{K}$ imply a radius of $R = 6.4R_{\odot}$ \citep{bell1995psr}. The
B star has a projected surface rotation velocity $v \sin i = 113 \pm
10\;\mathrm{km/s}$ \citep{bell1995psr}, consistent with rapid rotation (the
breakup velocity is $\sqrt{GM / R} = 512\;\mathrm{km/s}$). While there are
uncertainties in some of these parameters (e.g., the pulsar and companion masses
could be larger by about $10\%$; see \citealp{thorsett1999neutron}), we adopt
the above values in our calculation below for better comparison with previous
works.

\citet{lai1996} and \citet{lai1997} explained the observed orbital decay of the
PSR J0045 binary in terms of tidal excitations of discrete (rotation-modified)
g-modes followed by radiative damping. He showed that retrograde rotation of the
B-star \citep[consistent with the observed nodal precession of the binary orbit;
see][]{lai1995, kaspi1996params} can significantly enhance the strength of mode
excitation, and potentially account for the observed orbital decay, although he
did not compute the mode damping rate quantitatively.
\citet{kumar1997differential, kumar1998} examined the damping of g-modes and
showed that radiative damping in rigidly rotating B-star is inadequate to
explain the observed orbital decay rate, suggesting that in addition to
retrograde rotation, significant differential rotation or nonlinear parametric
mode decay may be required. Given the uncertainties in their estimates, it was
not clear to what extent the observed orbital decay rate can be quantitatively
explained. Moreover, both \citet{lai1995} and \citet{kumar1997differential,
kumar1998} considered discrete g-modes, whereas absorption of tidally excited
gravity waves near the stellar envelope, via either efficient radiative damping
or nonlinear breaking \citep[see][]{su2020}, implies outgoing waves (as adopted
in the works of \citealp{zahn1975dynamical, goldreich1989tidal, kushnir}) rather
than discrete g-modes.

We now consider the evolution of stellar spin and binary orbit for the PSR
J0045-7319 system. For the adopted system parameters above, the pericentre
distance is $a_{\rm p} \simeq 3.78R$, and pericentre frequency is
$\Omega_{\rm p}=\sqrt{(1+e)GM_{\rm tot}/a_{\rm p}^3}\simeq 0.20\sqrt{GM/R^3}$.
First we note that the tidal torque $T$ and the energy transfer rate
$\dot{E}_{\rm in}$ (dominated by the $m=2$ term) are related by
\begin{equation}
  {\dot E}_{\rm in}\simeq T \Omega_{\rm p} {f_2\over 5(1+e)^2 f_5},
\end{equation}
with $f_2=8.38$, $f_5=3.12$. Thus the spin evolution rate (for spin-orbit
synchronization and alignment) is given by
\begin{equation}
\left|{\dot {\bf S}\over S}\right|\simeq {L\over S} {5(1-e^2) f_5\over 2f_2}
\left|{\dot E_{\rm in}\over E_{\rm orb}}\right|,
\end{equation}
where $S=kMR^2 \bar\Omega_{\rm s}$ is the spin angular momentum of the star (with
$\bar\Omega_{\rm s}$ the mean rotation rate, and $k\simeq 0.1$), $L$ is the orbital
angular momentum, and $E_{\rm orb}$ is the orbital energy. Thus
\begin{equation}
\left|{\dot {\bf S}\over S}\right|\simeq 6.3\, {\Omega_{\rm p}\over
\bar\Omega_{\rm s}}
\left|{\dot a\over a}\right|.
\end{equation}
The observed nodal precession of the PSR J0045-7319 binary implies that the spin
of the B-star is far from spin-orbit alignment and synchronization driven by
tides \citep{lai1995}. Thus we require $\bar\Omega_{\rm s} \gtrsim
6.3\Omega_{\rm p}\sim \sqrt{GM/R^3}$, suggesting that the internal rotation rate
of the star is much larger that the surface rate.

Using Eq.~\eqref{eq:ein_dot_tot} with $e=0.808$ (and keeping only the $m=2$
term), we find that the orbital decay rate is given by
\begin{align}
\left|{\dot a\over a}\right|\simeq{}& 1.6\times 10^5\,\Omega\, {M_2\over M}
\left({M_{\rm tot}\over M_{\rm c}}\right)^{4/3}\nonumber\\
&\times \left({r_{\rm c}\over a}\right)^9{\rho_{\rm c}\over\bar\rho_{\rm c}}
\left(1-{\rho_{\rm c}\over\bar\rho_{\rm c}}\right)^2
\left|{4\over\gamma_E}
\left(1-{\gamma_E\Omega_{\rm s}\over\eta_2\Omega}\right)\right|^{8/3}.
\end{align}
With $\eta_2=10.55$, $a\simeq 19.7R$, and adopting $\rho_{\rm c}/\bar\rho_{\rm
c}\simeq
1/3$, we find the orbital decay rate
\begin{equation}
\left|{\dot P\over P}\right|\simeq {1\over 0.5\,{\rm Myr}}\,
\left({3M_\odot\over M_{\rm c}}\right)^{4/3}
\left({r_{\rm c}\over 0.54R}\right)^9
\left|1-{0.89\Omega_{\rm s}\over\Omega_{\rm p}}\right|^{8/3}.
\end{equation}
Note $\Omega_{\rm s}$ in the above expression refers to the rotation rate at the
radiative-convective boundary (RCB) $r_{\rm c}$, and we expect $\Omega_{\rm
s}\gtrsim \bar\Omega_{\rm s} \gtrsim 6\Omega_{\rm p}$.

\begin{figure}
    \centering
    \includegraphics[width=\colummwidth]{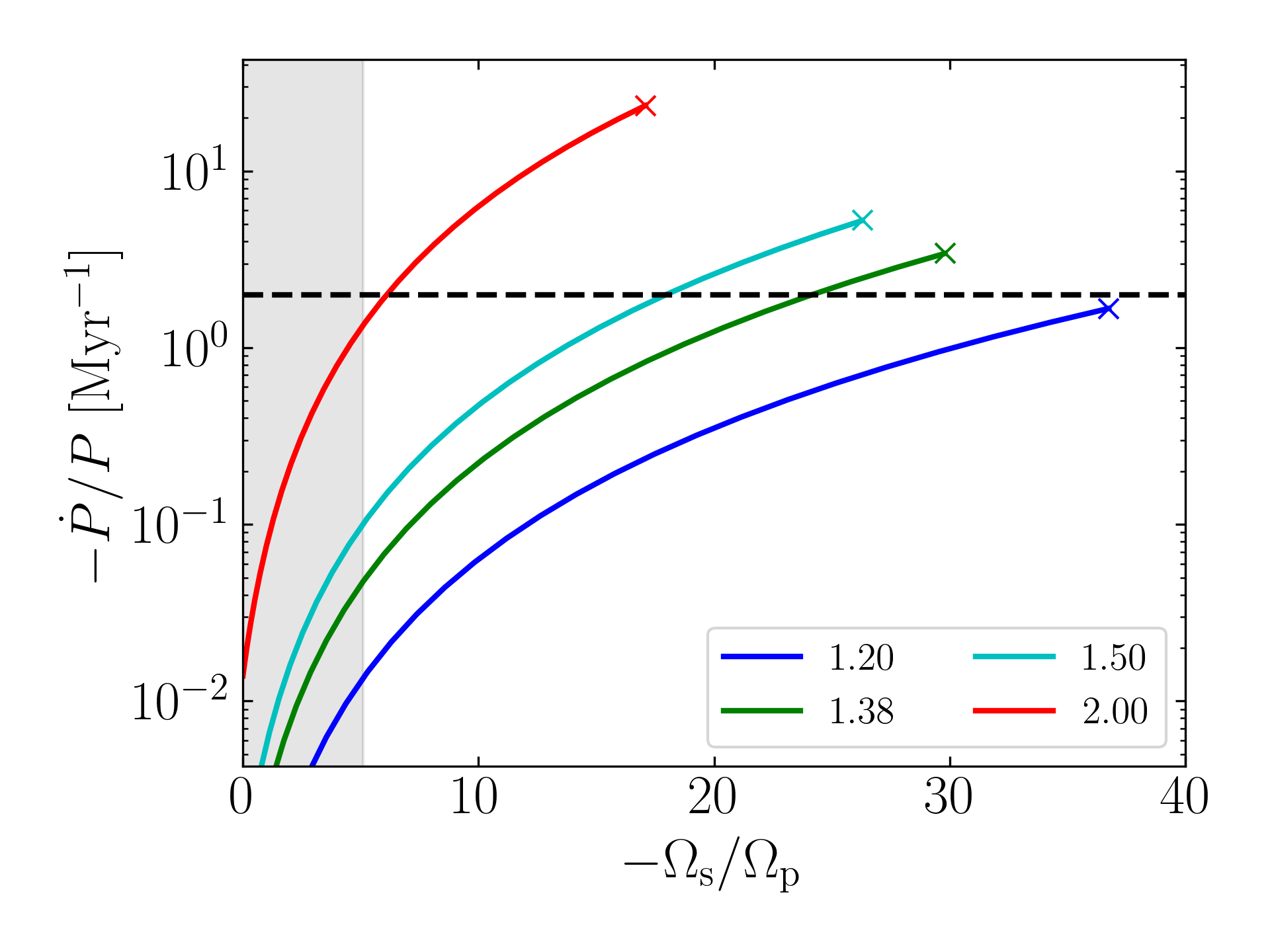}
    \caption{The orbital decay ratio $\dot{P} / P$ as a function of $\Omega_{\rm
    s}$ for the canonical parameters of the PSR J0045-7319 binary system, as
    evaluated by Eq.~\eqref{eq:ein_dot_tot}, for four different values of
    $r_{\rm c}$ (legend, in units of $R_{\odot}$). The measured $\dot{P} / P =
    -\p{0.5\;\mathrm{Myr}}^{-1}$ is shown by the horizontal dashed line. The
    vertical shaded region is the region where $\abs{\Omega_{\rm s}}$ is less
    than the breakup rotation rate of the star as a whole, given by $(GM /
    R^3)^{1/2}$. Each $r_{\rm c}$ is only shown for $\abs{\Omega_{\rm s}} \leq
    \Omega_{\rm s, c} \equiv (GM_{\rm c} / r_{\rm c}^3)^{1/2}$, the \emph{core}
    breakup rotation rate, and the colored crosses denote where $\Omega_{\rm s}
    = \Omega_{\rm s, c}$ for each value of $r_{\rm c}$. Note that for $r_{\rm c}
    \lesssim 1.3 R_{\odot}$, even a maximally rotating core cannot generate
    enough tidal torque to match the observed $\dot{P} /
    P$.}\label{fig:j0045_fid}
\end{figure}

To explain the observed orbital decay timescale $|P/\dot P|\simeq 0.5$~Myr, the
most critical parameter is the core radius $r_{\rm c}$. Kumar \& Quataert (1998)
adopted $M_{\rm c}\simeq 3M_{\odot}$ and $r_{\rm c}\simeq 1.38R_{\odot}$ (or
$r_{\rm c}\simeq 0.22R$) based on comparison with a Yale stellar evolution
model. This value of $r_{\rm c}$ would require $\Omega_{\rm s}/\Omega_{\rm
p}\simeq -24$, or $\Omega_{\rm s}\simeq -4.8 \sqrt{GM/R^3}$, to explain the
observed decay rate. If we use a slightly larger core radius, $r_{\rm
c}=1.5R_\odot \simeq 0.234 R$, the required $\Omega_{\rm s}$ would be
$\Omega_{\rm s}\simeq -18\Omega_{\rm p}\simeq -3.6\sqrt{GM/R^3}$.
In either case, extreme differential rotation (with $|\Omega_{\rm s}|$ at the
RCB at least a factor of few larger than the surface rotation) is required. In
Fig.~\ref{fig:j0045_fid} we show $\dot{P}/P$ as a function of $\Omega_{\rm s}$,
obtained using the exact expression Eq.~\eqref{eq:ein_dot_tot} for four
different values of $r_{\rm c}$. In general, given that the surface rotation
rate of the star is less than $\sqrt{GM/R^3}$, our calculation suggests that the
B-star must have large, retrograde differential rotation in order to explain the
observed orbital decay.

We note that our own exploration of stellar models using the Modules for
Experiments in Stellar Astrophysics \citep[MESA;][]{Paxton2011, Paxton2013,
Paxton2015, Paxton2018, Paxton2019} generally finds a small core radius ($r_{\rm
c}\lesssim R_\odot$) for the B-star in the PSR J0045-7319 system. Since the
system is in the Small Magellanic Cloud, we use a typical metallicity $Z =
0.1Z_{\odot}$, where $Z_{\odot}$ is the solar metallicity. For a range of
initial stellar masses, we begin with a non-rotating zero-age main sequence
(ZAMS) stellar model, and evolve it until its core hydrogen is depleted. Among
these evolutionary tracks, we select the model and stellar age that best matches
the observed $L$ and $T_{\rm surf}$ and record the core radius at that age. We
then repeat this procedure with different values for the convective overshoot
parameter, initial rotation rate (up to nearly maximally rotating), and
metallicity (up to $Z = 0.2Z_{\odot}$). For all parameter combinations, the
predicted stellar masses lie in the range $8.8M_{\odot} \leq M \leq
10M_{\odot}$, in agreement with the estimates in the literature. However, the
core radius for all stellar models is $r_{\rm c} \lesssim R_{\odot} \approx
0.16R$. This reflects the fact that our best-fitting stellar models tend to be
somewhat evolved from the ZAMS\@.

We caution that applying standard isolated stellar evolution to the B-star in
the PSR J0045-7319 system can be fraught with uncertainties. The very rapid
rotation may induce instabilities and allow the star to transport material from
the hydrogen-rich envelope into the central burning region and vice versa,
potentially making the convective core larger \citep[e.g.][]{maeder1987,
heger2000}. This effect may be enhanced by the large misaligned, differential
rotation. Related to the rotational effect is the mass transfer effect: The
observed rapid orbital decay rate and general orbital evolution modeling sets an
upper limit of about $1.4$~Myr on the age of the binary system since the last
supernova explosion \citep{lai1996}, much shorter than the canonical MS
lifetime. Thus the B-star must have accreted significant material from the
pulsar progenitor in the recent past. Such accretion can affect the structure
and evolution of the B-star in a significant way. Given these uncertainties, we
think it is likely that the B-star has a larger core radius than the canonical
value, rendering a less extreme differential rotation\footnote{We note that a
recent study of intermediate and high-mass eclipsing binaries suggests that
convective core masses are underpredicted by stellar structure codes
\citep{larger_conv_masses}.}.

\section{Summary and Discussion}\label{s:disc}

\subsection{Key Results}

The main goal of this paper is to derive easy-to-use, analytical expressions for
the tidal torque $T$ and tidal energy transfer rate $\dot E_{\rm in}$ due to
internal gravity wave (IGW) dissipation in a massive, main-sequence (MS) star
under the gravitational influence of an eccentric companion. Tidal evolution in
such systems plays an important role in the formation scenarios of merging
neutron-star (NS) binaries. For general eccentricities, these expressions are
given by the sums, Eqs.~\eqref{eq:tau_explicit_sum}
and~\eqref{eq:ein_explicit_sum}, respectively. However, when the eccentricity is
large, these sums require the evaluation of many terms to be accurate. We have
derived approximate expressions in two regimes:
\begin{itemize}
    \item For $e \gtrsim 0.3$, we show that the tidal torque and energy transfer
        rate can be accurately approximated by the integral expressions,
        Eqs.~\eqref{eq:tau_int} and~\eqref{eq:ein_int}.
    \item If furthermore the spin of the stellar core is not very close to its
        pseudo-synchronized value (i.e.\ where the torque vanishes; see
        Section~\ref{ss:}), we show that these two integral
        expressions can be well approximated by the closed-form expressions,
        Eqs.~\eqref{eq:tau_approx} and~\eqref{eq:ein_dot_tot}.
\end{itemize}
These analytical expressions for $T$ and $\dot E_{\rm in}$ (particularly the
closed-form expressions) can be easily applied to study the spin and orbital
evolution of eccentric binaries consisting of massive MS stars, such as those
found in the evolution scenarios leading to the formation of merging neutron
star binaries.

We then apply our analytical expressions to the PSR J0045-7319 binary system,
which has a massive B-star companion and a measured orbital decay rate. This
system provides a rare example to test/calibrate our theoretical understanding
of IGW-mediated dynamical tides in massive stars. We show that for the
``standard'' radius of the convective core based on isolated non-rotating
stellar modeling, the B-star must have a significant retrograde and differential
rotation (with the rotation rate at the convective-radiative boundary at least a
few times larger than the surface rotation rate of the star). Alternatively, we
suggest that the convective core radius may be larger than the standard value as
a result of rapid stellar rotation and/or mass transfer to the B-star in the
recent past during the post-MS evolution of the pulsar progenitor. Overall, our
attempt to explain the rapid orbital decay of the PSR J0045-7319 binary
highlights the critical importance of internal stellar structure, particularly
the size of the convective core, in determining the tidal evolution of eccentric
binaries containing massive MS stars.

\subsection{Caveats}

Concerning our analytical expressions for the tidal torque and energy transfer
rate, we note a few potential caveats:

(i) For simplicity, we have assumed that the stellar spin and orbit axes are
aligned or anti-aligned. For general stellar obliquities (as in the case of the
PSR J0045-7319 binary system), it is possible to decompose the tidal force into
different Fourier components (see Appendix A of \citealp{lai1997}), and this
would unlikely to yield qualitatively different result in terms of the orbital
decay rate \citep[cf.][]{lai2012tidal}.

(ii) More importantly, in our work we have assumed that all Fourier harmonics
(with different forcing frequencies $N\Omega$ in the inertial frame, where
$\Omega$ is the mean orbital frequency) of the tidal potential excite IGWs at
the radiative-convective boundary that damp completely as they propagate towards
the stellar surface. This is known to be the case when the normalized tidal
forcing frequency $\omega$ (equal to $N\Omega-2\Omega_{\rm s}$ for $m=2$ and
$N\Omega$ for $m=0$) satisfies $|\omega| \ll \sqrt{GM / R^3}$ due to efficient
radiative damping (\citealp{zahn1975dynamical, kushnir}; see also the review
paper \citealp{ogilvie2014tidal}). Note that the dominant Fourier harmonics of
the tidal potential have $N\Omega\sim\Omega_{\rm p}$ (the orbital frequency at
the pericentre). Thus, as long as $\Omega_{\rm p}, |\Omega_{\rm s}| \ll
\sqrt{GM/R^3}$, the assumption of outgoing IGWs inherent in our theory is valid.
When $|\omega|$ is comparable to $\sqrt{GM/R^3}$, it is traditionally thought
that IGWs set up standing modes in the stellar interior. This is because IGWs
are evanescent sufficiently near stellar surface, where the pressure scale
height becomes smaller than the radial wavelength of the wave
\citep{goldreich1989tidal}. Moreover, \citet{su2020} found that sufficiently
large-amplitude IGWs can spontaneously cause a critical layer to form due to
nonlinear wave breaking, causing incident IGWs to be efficiently absorbed. After
such a critical layer forms, it can propagate to deep within the stellar
interior and efficiently absorb IGWs even before they reach breaking amplitudes.
Wave breaking and other nonlinear effects (such as nonlinear mode couplings) may
be enhanced in an eccentric binary such as PSR J0045-7319, where IGWs with a
wide range of frequencies are excited. \citet{burkart2012tidal} found that
the frequency cutoff for g-modes in the eccentric heartbeat star system KOI-54
(with eccentricity $e = 0.83$, orbital period $P = 41.8\;\mathrm{days}$, and
stellar mass $M \approx 2.3M_{\odot}$) is in agreement with the frequency at
which IGWs become efficiently damped by thermal diffusion. However, in
higher-mass MS stars such as that in PSR J0045-7319, nonlinear wave breaking
may be an important source of IGW damping due to the larger tidal torque. More
work is needed to address whether our assumption of traveling IGWs in massive
stars is accurate.

\section{Acknowledgements}

We thank Matteo Cantiello, Christopher O'Connor, Eliot Quataert, and Michelle
Vick for fruitful discussions. YS is supported by the NASA FINESST grant
19-ASTRO19-0041. This work has been supported in part by NSF grants AST-1715246
and AST-2107796.

\section{Data Availability}

The data underlying this article will be shared on reasonable request to the
corresponding author.

\bibliographystyle{mnras}
\bibliography{Su_eccentric_tides}

\begin{thebibliography}{}
\makeatletter
\relax
\def\mn@urlcharsother{\let\do\@makeother \do\$\do\&\do\#\do\^\do\_\do\%\do\~}
\def\mn@doi{\begingroup\mn@urlcharsother \@ifnextchar [ {\mn@doi@}
  {\mn@doi@[]}}
\def\mn@doi@[#1]#2{\def\@tempa{#1}\ifx\@tempa\@empty \href
  {http://dx.doi.org/#2} {doi:#2}\else \href {http://dx.doi.org/#2} {#1}\fi
  \endgroup}
\def\mn@eprint#1#2{\mn@eprint@#1:#2::\@nil}
\def\mn@eprint@arXiv#1{\href {http://arxiv.org/abs/#1} {{\tt arXiv:#1}}}
\def\mn@eprint@dblp#1{\href {http://dblp.uni-trier.de/rec/bibtex/#1.xml}
  {dblp:#1}}
\def\mn@eprint@#1:#2:#3:#4\@nil{\def\@tempa {#1}\def\@tempb {#2}\def\@tempc
  {#3}\ifx \@tempc \@empty \let \@tempc \@tempb \let \@tempb \@tempa \fi \ifx
  \@tempb \@empty \def\@tempb {arXiv}\fi \@ifundefined
  {mn@eprint@\@tempb}{\@tempb:\@tempc}{\expandafter \expandafter \csname
  mn@eprint@\@tempb\endcsname \expandafter{\@tempc}}}

\bibitem[\protect\citeauthoryear{Alexander}{Alexander}{1973}]{alexander73}
Alexander M.,  1973, Astrophysics and Space Science, 23, 459

\bibitem[\protect\citeauthoryear{Bell, Bessell, Stappers, Bailes  \&
  Kaspi}{Bell et~al.}{1995}]{bell1995psr}
Bell J.,  Bessell M.,  Stappers B.,  Bailes M.,   Kaspi V.,  1995, The
  Astrophysical Journal Letters, 447, L117

\bibitem[\protect\citeauthoryear{Burkart, Quataert, Arras  \& Weinberg}{Burkart
  et~al.}{2012}]{burkart2012tidal}
Burkart J.,  Quataert E.,  Arras P.,   Weinberg N.~N.,  2012, Monthly Notices
  of the Royal Astronomical Society, 421, 983

\bibitem[\protect\citeauthoryear{Champion et~al.,}{Champion
  et~al.}{2008}]{champion2008eccentric}
Champion D.~J.,  et~al., 2008, Science, 320, 1309

\bibitem[\protect\citeauthoryear{Correia, Bou{\'e}, Laskar  \&
  Rodr{\'\i}guez}{Correia et~al.}{2014}]{correia2014deformation}
Correia A.~C.,  Bou{\'e} G.,  Laskar J.,   Rodr{\'\i}guez A.,  2014, Astronomy
  \& Astrophysics, 571, A50

\bibitem[\protect\citeauthoryear{Goldreich \& Nicholson}{Goldreich \&
  Nicholson}{1989}]{goldreich1989tidal}
Goldreich P.,  Nicholson P.~D.,  1989, Astrophysical Journal, 342, 1079

\bibitem[\protect\citeauthoryear{{Heger} \& {Langer}}{{Heger} \&
  {Langer}}{2000}]{heger2000}
{Heger} A.,  {Langer} N.,  2000, \mn@doi [\apj] {10.1086/317239}, \href
  {https://ui.adsabs.harvard.edu/abs/2000ApJ...544.1016H} {544, 1016}

\bibitem[\protect\citeauthoryear{Hurley, Tout  \& Pols}{Hurley
  et~al.}{2002}]{hurley2002evolution}
Hurley J.~R.,  Tout C.~A.,   Pols O.~R.,  2002, Monthly Notices of the Royal
  Astronomical Society, 329, 897

\bibitem[\protect\citeauthoryear{Hut}{Hut}{1981}]{hut81}
Hut P.,  1981, Astronomy and Astrophysics, 99, 126

\bibitem[\protect\citeauthoryear{Janka, Wongwathanarat, Kramer  et~al.}{Janka
  et~al.}{2021}]{janka2021supernova}
Janka H.,  Wongwathanarat A.,  Kramer M.,   et~al., 2021, arXiv preprint
  arXiv:2104.07493

\bibitem[\protect\citeauthoryear{{Johnston}, {Manchester}, {Lyne}, {Bailes},
  {Kaspi}, {Qiao}  \& {D'Amico}}{{Johnston} et~al.}{1992}]{nsms1}
{Johnston} S.,  {Manchester} R.~N.,  {Lyne} A.~G.,  {Bailes} M.,  {Kaspi}
  V.~M.,  {Qiao} G.,   {D'Amico} N.,  1992, \mn@doi [\apjl] {10.1086/186300},
  \href {https://ui.adsabs.harvard.edu/abs/1992ApJ...387L..37J} {387, L37}

\bibitem[\protect\citeauthoryear{Johnston, Manchester, Lyne, Nicastro  \&
  Spyromilio}{Johnston et~al.}{1994}]{johnston1994radio}
Johnston S.,  Manchester R.,  Lyne A.,  Nicastro L.,   Spyromilio J.,  1994,
  Monthly Notices of the Royal Astronomical Society, 268, 430

\bibitem[\protect\citeauthoryear{Kaspi, Johnston, Bell, Manchester, Bailes,
  Bessell, Lyne  \& D'amico}{Kaspi et~al.}{1994}]{kaspi1994massive}
Kaspi V.,  Johnston S.,  Bell J.,  Manchester R.,  Bailes M.,  Bessell M.,
  Lyne A.~a.,   D'amico N.,  1994, The Astrophysical Journal, 423, L43

\bibitem[\protect\citeauthoryear{Kaspi, Bailes, Manchester, Stappers  \&
  Bell}{Kaspi et~al.}{1996}]{kaspi1996params}
Kaspi V.,  Bailes M.,  Manchester R.,  Stappers B.,   Bell J.,  1996, Nature,
  381, 584

\bibitem[\protect\citeauthoryear{Kumar \& Quataert}{Kumar \&
  Quataert}{1997}]{kumar1997differential}
Kumar P.,  Quataert E.~J.,  1997, The Astrophysical Journal Letters, 479, L51

\bibitem[\protect\citeauthoryear{Kumar \& Quataert}{Kumar \&
  Quataert}{1998}]{kumar1998}
Kumar P.,  Quataert E.~J.,  1998, The Astrophysical Journal, 493, 412

\bibitem[\protect\citeauthoryear{Kushnir, Zaldarriaga, Kollmeier  \&
  Waldman}{Kushnir et~al.}{2017}]{kushnir}
Kushnir D.,  Zaldarriaga M.,  Kollmeier J.~A.,   Waldman R.,  2017, Monthly
  Notices of the Royal Astronomical Society, 467, 2146

\bibitem[\protect\citeauthoryear{Lai}{Lai}{1996}]{lai1996}
Lai D.,  1996, The Astrophysical Journal Letters, 466, L35

\bibitem[\protect\citeauthoryear{{Lai}}{{Lai}}{1997}]{lai1997}
{Lai} D.,  1997, \mn@doi [\apj] {10.1086/304899}, \href
  {https://ui.adsabs.harvard.edu/abs/1997ApJ...490..847L} {490, 847}

\bibitem[\protect\citeauthoryear{Lai}{Lai}{2012}]{lai2012tidal}
Lai D.,  2012, Monthly Notices of the Royal Astronomical Society, 423, 486

\bibitem[\protect\citeauthoryear{{Lai}, {Bildsten}  \& {Kaspi}}{{Lai}
  et~al.}{1995}]{lai1995}
{Lai} D.,  {Bildsten} L.,   {Kaspi} V.~M.,  1995, \mn@doi [\apj]
  {10.1086/176350}, \href
  {https://ui.adsabs.harvard.edu/abs/1995ApJ...452..819L} {452, 819}

\bibitem[\protect\citeauthoryear{Lai, Chernoff  \& Cordes}{Lai
  et~al.}{2001}]{lai2001pulsar}
Lai D.,  Chernoff D.~F.,   Cordes J.~M.,  2001, The Astrophysical Journal, 549,
  1111

\bibitem[\protect\citeauthoryear{{Lorimer} et~al.,}{{Lorimer}
  et~al.}{2006}]{nsms2}
{Lorimer} D.~R.,  et~al., 2006, \mn@doi [\mnras]
  {10.1111/j.1365-2966.2006.10887.x}, \href
  {https://ui.adsabs.harvard.edu/abs/2006MNRAS.372..777L} {372, 777}

\bibitem[\protect\citeauthoryear{{Lyne}, {Stappers}, {Keith}, {Ray}, {Kerr},
  {Camilo}  \& {Johnson}}{{Lyne} et~al.}{2015}]{nsms4}
{Lyne} A.~G.,  {Stappers} B.~W.,  {Keith} M.~J.,  {Ray} P.~S.,  {Kerr} M.,
  {Camilo} F.,   {Johnson} T.~J.,  2015, \mn@doi [\mnras]
  {10.1093/mnras/stv236}, \href
  {https://ui.adsabs.harvard.edu/abs/2015MNRAS.451..581L} {451, 581}

\bibitem[\protect\citeauthoryear{{Maeder}}{{Maeder}}{1987}]{maeder1987}
{Maeder} A.,  1987, \aap, \href
  {https://ui.adsabs.harvard.edu/abs/1987A&A...178..159M} {178, 159}

\bibitem[\protect\citeauthoryear{Murray \& Dermott}{Murray \&
  Dermott}{1999}]{murray1999solar}
Murray C.~D.,  Dermott S.~F.,  1999, Solar system dynamics.
Cambridge university press

\bibitem[\protect\citeauthoryear{Ogilvie}{Ogilvie}{2014}]{ogilvie2014tidal}
Ogilvie G.~I.,  2014, Annual Review of Astronomy and Astrophysics, 52, 171

\bibitem[\protect\citeauthoryear{O'Leary \& Burkart}{O'Leary \&
  Burkart}{2014}]{oleary}
O'Leary R.~M.,  Burkart J.,  2014, \mn@doi [Monthly Notices of the Royal
  Astronomical Society] {10.1093/mnras/stu335}, 440, 3036

\bibitem[\protect\citeauthoryear{{Paxton}, {Bildsten}, {Dotter}, {Herwig},
  {Lesaffre}  \& {Timmes}}{{Paxton} et~al.}{2011}]{Paxton2011}
{Paxton} B.,  {Bildsten} L.,  {Dotter} A.,  {Herwig} F.,  {Lesaffre} P.,
  {Timmes} F.,  2011, \mn@doi [\apjs] {10.1088/0067-0049/192/1/3}, \href
  {https://ui.adsabs.harvard.edu/abs/2011ApJS..192....3P} {192, 3}

\bibitem[\protect\citeauthoryear{{Paxton} et~al.,}{{Paxton}
  et~al.}{2013}]{Paxton2013}
{Paxton} B.,  et~al., 2013, \mn@doi [\apjs] {10.1088/0067-0049/208/1/4}, \href
  {https://ui.adsabs.harvard.edu/abs/2013ApJS..208....4P} {208, 4}

\bibitem[\protect\citeauthoryear{{Paxton} et~al.,}{{Paxton}
  et~al.}{2015}]{Paxton2015}
{Paxton} B.,  et~al., 2015, \mn@doi [\apjs] {10.1088/0067-0049/220/1/15}, \href
  {https://ui.adsabs.harvard.edu/abs/2015ApJS..220...15P} {220, 15}

\bibitem[\protect\citeauthoryear{{Paxton} et~al.,}{{Paxton}
  et~al.}{2018}]{Paxton2018}
{Paxton} B.,  et~al., 2018, \mn@doi [\apjs] {10.3847/1538-4365/aaa5a8}, \href
  {https://ui.adsabs.harvard.edu/abs/2018ApJS..234...34P} {234, 34}

\bibitem[\protect\citeauthoryear{{Paxton} et~al.,}{{Paxton}
  et~al.}{2019}]{Paxton2019}
{Paxton} B.,  et~al., 2019, \mn@doi [\apjs] {10.3847/1538-4365/ab2241}, \href
  {https://ui.adsabs.harvard.edu/abs/2019ApJS..243...10P} {243, 10}

\bibitem[\protect\citeauthoryear{Savonije \& Papaloizou}{Savonije \&
  Papaloizou}{1983}]{savonije1983tidal}
Savonije G.,  Papaloizou J.,  1983, Monthly Notices of the Royal Astronomical
  Society, 203, 581

\bibitem[\protect\citeauthoryear{{Stairs} et~al.,}{{Stairs}
  et~al.}{2001}]{nsms3}
{Stairs} I.~H.,  et~al., 2001, \mn@doi [\mnras]
  {10.1046/j.1365-8711.2001.04447.x}, \href
  {https://ui.adsabs.harvard.edu/abs/2001MNRAS.325..979S} {325, 979}

\bibitem[\protect\citeauthoryear{Stein \& Shakarchi}{Stein \&
  Shakarchi}{2009}]{stein2009real}
Stein E.~M.,  Shakarchi R.,  2009, Real analysis: measure theory, integration,
  and Hilbert spaces.
Princeton University Press

\bibitem[\protect\citeauthoryear{Storch \& Lai}{Storch \& Lai}{2013}]{sl}
Storch N.~I.,  Lai D.,  2013, Monthly Notices of the Royal Astronomical
  Society, 438, 1526

\bibitem[\protect\citeauthoryear{Su, Lecoanet  \& Lai}{Su
  et~al.}{2020}]{su2020}
Su Y.,  Lecoanet D.,   Lai D.,  2020, \mn@doi [Monthly Notices of the Royal
  Astronomical Society] {10.1093/mnras/staa1306}, 495, 1239

\bibitem[\protect\citeauthoryear{{Tauris} et~al.,}{{Tauris}
  et~al.}{2017}]{tauris2017}
{Tauris} T.~M.,  et~al., 2017, \mn@doi [\apj] {10.3847/1538-4357/aa7e89}, \href
  {https://ui.adsabs.harvard.edu/abs/2017ApJ...846..170T} {846, 170}

\bibitem[\protect\citeauthoryear{Thorsett \& Chakrabarty}{Thorsett \&
  Chakrabarty}{1999}]{thorsett1999neutron}
Thorsett S.~E.,  Chakrabarty D.,  1999, The Astrophysical Journal, 512, 288

\bibitem[\protect\citeauthoryear{Tkachenko et~al.,}{Tkachenko
  et~al.}{2020}]{larger_conv_masses}
Tkachenko A.,  et~al., 2020, Astronomy \& Astrophysics, 637, A60

\bibitem[\protect\citeauthoryear{Vick, Lai  \& Fuller}{Vick et~al.}{2017}]{vlf}
Vick M.,  Lai D.,   Fuller J.,  2017, Monthly Notices of the Royal Astronomical
  Society, 468, 2296

\bibitem[\protect\citeauthoryear{{Vick}, {MacLeod}, {Lai}  \& {Loeb}}{{Vick}
  et~al.}{2021}]{vicklai2021}
{Vick} M.,  {MacLeod} M.,  {Lai} D.,   {Loeb} A.,  2021, \mn@doi [\mnras]
  {10.1093/mnras/stab850}, \href
  {https://ui.adsabs.harvard.edu/abs/2021MNRAS.503.5569V} {503, 5569}

\bibitem[\protect\citeauthoryear{Vigna-G{\'o}mez, MacLeod, Neijssel,
  Broekgaarden, Justham, Howitt, de Mink  \& Mandel}{Vigna-G{\'o}mez
  et~al.}{2020}]{vigna2020common}
Vigna-G{\'o}mez A.,  MacLeod M.,  Neijssel C.~J.,  Broekgaarden F.~S.,  Justham
  S.,  Howitt G.,  de Mink S.~E.,   Mandel I.,  2020, arXiv preprint
  arXiv:2001.09829

\bibitem[\protect\citeauthoryear{Yu, Weinberg  \& Fuller}{Yu
  et~al.}{2020}]{yu2020non}
Yu H.,  Weinberg N.~N.,   Fuller J.,  2020, Monthly Notices of the Royal
  Astronomical Society, 496, 5482

\bibitem[\protect\citeauthoryear{Yu, Fuller  \& Burdge}{Yu
  et~al.}{2021}]{yu2021tidally}
Yu H.,  Fuller J.,   Burdge K.~B.,  2021, Monthly Notices of the Royal
  Astronomical Society, 501, 1836

\bibitem[\protect\citeauthoryear{Zahn}{Zahn}{1975}]{zahn1975dynamical}
Zahn J.-P.,  1975, Astronomy and Astrophysics, 41, 329

\bibitem[\protect\citeauthoryear{Zaldarriaga, Kushnir  \&
  Kollmeier}{Zaldarriaga et~al.}{2018}]{zaldarriaga2018expected}
Zaldarriaga M.,  Kushnir D.,   Kollmeier J.~A.,  2018, Monthly Notices of the
  Royal Astronomical Society, 473, 4174

\makeatother
\end{thebibliography}


\bsp
\label{lastpage} 
\end{document}